\documentclass{emulateapj}
\pdfoutput=1
\usepackage{apjfonts,amsmath}
\usepackage{graphicx}
\newcommand{\beq}{\begin{equation}}
\newcommand{\eeq}{\end{equation}}
\newcommand{\bea}{\begin{eqnarray}}
\newcommand{\eea}{\end{eqnarray}}

\begin{document}

\title{Collapsar Accretion and the Gamma-Ray Burst X-Ray Light Curve}
\author{
Christopher C. Lindner
\altaffilmark{1},
Milo\v s Milosavljevi\'c
\altaffilmark{1,2},
Sean M. Couch
\altaffilmark{1},
and
Pawan Kumar
\altaffilmark{1} 
}
\altaffiltext{1}{Department of Astronomy, University of Texas, 1 University Station C1400, Austin, TX 78712.}
\altaffiltext{2}{Texas Cosmology Center, University of Texas, 1 University Station C1400, Austin, TX 78712.}

\righthead{COLLAPSAR ACCRETION AND X-RAY LIGHT CURVE}
\lefthead{LINDNER ET AL.}

\begin{abstract}

We present axisymmetric hydrodynamical simulations of the long-term accretion of a rotating gamma-ray burst progenitor star, a ``collapsar,'' onto the central compact object, which we take to be a black hole.  The simulations were carried out with the adaptive mesh refinement code FLASH in two spatial dimensions and with an explicit shear viscosity.  The evolution of the central accretion rate exhibits phases reminiscent of the long GRB $\gamma$-ray and X-ray light curve, which lends support to the proposal by \citet{Kumar:08a,Kumar:08b} that the luminosity is modulated by the central accretion rate.  In the first ``prompt'' phase characterized by an approximately constant accretion rate, the black hole acquires most of its final mass through supersonic quasiradial accretion occurring at a steady rate of $\sim 0.2~M_\odot ~\textrm{s}^{-1}$.  After a few tens of seconds, an accretion shock sweeps outward through the star.  The formation and outward expansion of the accretion shock is accompanied with a sudden and rapid power-law decline in the central accretion rate $\dot M\propto t^{-2.8}$, which resembles the $L_{\rm X}\propto t^{-3}$ decline observed in the X-ray light curves.  The collapsed, shock-heated stellar envelope settles into a thick, low-mass equatorial disk embedded within a massive, pressure-supported atmosphere, similar to the picture proposed by \citet{Begelman:08} for ``quasistars.''  After a few hundred seconds, the inflow of low-angular-momentum material in the axial funnel reverses into an outflow from the surface of the thick disk. Meanwhile, the rapid decline of the accretion rate slows down, or even settles a in steady state with $\dot M\sim 5\times10^{-5}~M_\odot~\textrm{s}^{-1}$, which resembles the ``plateau'' phase in the X-ray light curve.  While the duration of the ``prompt'' phase depends on the resolution in our simulations, we provide an analytical model taking into account neutrino losses that estimates the duration to be $\sim 20~\textrm{s}$.  The model suggests that the steep decline in GRB X-ray light curves is triggered by the circularization of the infalling stellar envelope at radii where the virial temperature is below $10^{10} ~\textrm{K}$, such that neutrino cooling shuts off and an outward expansion of the accretion shock becomes imminent; GRBs with longer prompt $\gamma$-ray emission have more slowly rotating envelopes.


\end{abstract}

\section{Introduction}
\label{sec:intro}
\setcounter{footnote}{0}

Observations of long gamma-ray bursts (GRBs) carried out with the NASA {\it Swift} satellite have shown that the $\gamma$-ray prompt emission ceases after about a minute in the observer frame, corresponding to tens of seconds in the rest frame of the progenitor star.  The $\gamma$-ray light curve, converted to a fiducial X-ray spectral band, smoothly joins the X-ray light curve, which declines rapidly, ($t^{-3}$ or faster) lasting for about $80$ to $300~\textrm{s}$ \citep{Tagliaferri:05,Nousek:06,OBrien:06}.  The rapid decline is often followed by a phase, from about $\sim 10^3$ to $10^4~\textrm{s}$, during which the X-ray flux is roughly constant or declines more slowly with time.  The X-ray light curves of some GRBs exhibit ``flares'' where the flux increases suddenly by a factor of $\lesssim 10^2$ and drops precipitously, with the rise and decline associated with the flare occurring on a time scale much shorter than the age of the burst \citep[see, e.g.,][]{Burrows:05,Falcone:06}.  Following about $\sim10^3-10^4~\textrm{s}$, a more rapid decline of the luminosity resumes \citep[see, e.g.,][and references therein]{ZhangB:06}, and occasionally steepens further at $\sim10^4-10^5~\textrm{s}$ \citep[e.g.,][]{Vaughan:06}. 

The goal of the present work is to utilize two-dimensional hydrodynamic simulations to test the hypothesis \citep{Kumar:08a,Kumar:08b} that this characteristic structure of the X-ray light curve, which was summarized by \citet{ZhangB:06}, reflects a modulation in the rate of central accretion of a rotating progenitor star onto a black hole or a neutron star, as in the collapsar model of GRBs \citep{Woosley:93,MacFadyen:99,MacFadyen:01,Woosley:06b}.  We do not attempt to explore the implications of the potential presence of a magnetosphere, as in the magnetar model for GRBs \citep[e.g.,][]{Duncan:92,Wheeler:00,ZhangB:01,Thompson:04,Komissarov:07,Bucciantini:07,Bucciantini:09}. We will attempt to gain insight in the origin of the steady $\gamma$-ray luminosity (the prompt phase which we will refer to as ``Phase 0''), the rapid decline in the X-ray light curve (Phase I in the nomenclature of \citealt{ZhangB:06}), and phase of quasi-steady luminosity or slow decline (Phase II).  We will briefly attempt to extrapolate the results of our simulations to the subsequent steeper decline phases (Phases III and IV).  

\citet{Kumar:08a,Kumar:08b} obtained the key features of the $\gamma$-ray and X-ray light curve by estimating central accretion rate resulting from the free (i.e., ballistic) infall of a rotating progenitor star.  In this picture, the material that has sufficient initial angular momentum to circularize outside of the innermost stable circular orbit (ISCO) of the black hole, forms a disk in the equatorial plane, and subsequently accretes via disk accretion \citep{Narayan:01}.  If the luminosity is then assumed to be proportional to the central accretion rate, and if the distance of the $\gamma$-ray or X-ray emitting region from the center of the star is assumed to be approximately independent of time on time scales $10-10^5~\textrm{s}$, an accretion model directly translates into a synthetic light curve that can be compared with an observed light curve.  \citet{Kumar:08a,Kumar:08b} have shown that with the simplest accretion model involving ballistic infall onto the midplane (assumption also made by \citealt{Janiuk:08} and \citealt{Cannizzo:09}) and subsequent disk accretion, the mapping of the mass accretion history onto the light curve provides a powerful insight into the stratification and angular momentum structure of the progenitor star.  

In their ballistic infall model, \citet{Kumar:08a,Kumar:08b} were not able to discriminate between models in which the quasi-steady activity in Phase II arose from disk accretion, or from late-time accretion from an extended stellar envelope. Departures from ballistic infall are expected if the infalling material passes through an accretion shock \citep[see, e.g.,][]{MacFadyen:99,Lee:06,Nagataki:07,LopezCamara:09}, or if the disk launches a thermal \citep[][]{MacFadyen:03a,MacFadyen:03b,Kohri:05} or magnetohydrodynamic \citep[e.g.,][]{Proga:03c} outflow (``wind'') that can interfere with the infall.  The existence of the outflow is particularly interesting because of the potential for nucleosynthesis in the free neutron-rich outflow launched from the inner part of the disk \citep[see, e.g.,][]{Pruet:03,Surman:06,Fujimoto:07,Nagataki:07,Maeda:09} and because of the potential that the outflow can deplete the accreting stellar envelope and limit the envelope mass that is accreted onto the central black hole.

During the first $\sim10^2~\textrm{s}$ following the formation of the central black hole when the accretion rate is $\dot M\gg 10^{-3}M_\odot~\textrm{s}^{-1}$ (the precise condition depends on the black hole spin and shear stress-to-pressure ratio $\alpha$ in the disk), the inner accretion disk cools by neutrino emission and nuclear photodisintegration and accretes in a radiatively efficient fashion, except for in the very inner, optically thick region \citep[e.g.,][]{Popham:99,Narayan:01,DiMatteo:02,Chen:07}.  Instabilities in the thin disk have been cited as a candidate class of mechanisms that could produce the observed X-ray flares \citep{Perna:06,Lazzati:07,Lazzati:08} and could also produce detectable gravitational radiation \citep{Piro:07}.  Our global axisymmetric models are the necessary stepping stone toward the substantially more computationally demanding three-dimensional simulations that will be required to pin down any nonaxisymmetric instabilities in the accreting collapsar \citep[see, e.g.,][]{Rockefeller:06}.

We employ two-dimensional unmagnetized hydrodynamic simulations of the collapse, circularization, and accretion of a stellar envelope onto a central point mass, which we assume to be a black hole; relativistic corrections to the gravitational potential are ignored in our simulations since the innermost grid point lies at over $20$ Schwarzschild radii in the simulation extending to the smallest radius from the black hole.  The torque and dissipation arising from the $R-\phi$ component of the magnetic stress is emulated with a Navier-Stokes term parameterized by an $\alpha$-viscosity prescription. For comparison with the X-ray light curve, we measure the central accretion rate.  We track the flow of mass and energy at spherical radii $10^8~\textrm{cm}\lesssim r\lesssim  10^{11}~\textrm{cm}$ and interpret the results in view of the existing knowledge on radiatively-inefficient accretion flows.  We observe an outflow and measure the rate at which the accreting stellar envelope is lost to the outflow.   The mechanics of post-core-collapse accretion and outflows is key to estimating the final mass of the black hole and the nucleosynthetic composition of the ejected matter \citep[e.g.,][and references therein]{Zhang:08}.  The method that we develop here can in future be utilized to estimate the masses of the black holes resulting from the collapse of massive, initially metal-poor ``Population III'' stars as well as from the collapse of the even more massive, hypothetical ``supermassive stars,'' in the presence of rotation.

In this work we do not simulate the neutrino-cooled disk, and in the simulations simply impose that the mass that crosses the innermost cylindrical radius of our simulation, $R_{\rm min} = (0.5-2) \times 10^{8} ~\textrm{cm}$, is instantaneously incorporated inside the black hole and does not provide any further energetic feedback while at radii $R<R_{\rm min}$.  This very rough assumption is bound to fail in general; it is most compatible with the regime in which the transition from efficient to inefficient cooling occurs at $R\gtrsim R_{\rm min}$.  Since the transitional radius for efficient neutrino cooling recedes inward with the increasing stress-to-pressure ratio $\alpha$ for a given accretion rate \citep{Chen:07}, the assumption that cooling is efficient within $\lesssim 10^8~\textrm{ cm}$ is valid for $\alpha\lesssim 0.01$.   We also ignore nuclear photodisintegration when temperature rises above $\sim (5-10)\times10^9~\textrm{K}$; in reality, the photodissociation allows for some heating via the capture by free nucleons of the neutrinos emitted in the inner disk \citep{Nagataki:07}, which we do not model.  However, we do incorporate neutrino cooling in a simple analytical model for the evolution of the accretion shock at the radii that we do not resolve, $\lesssim 5\times10^7~\textrm{cm}$.  In combination with the simulations, the model provides a theory for the duration of the prompt emission phase observed in the $\gamma$-rays.

\citet{Cannizzo:09} speculate that a cool, thin disk may form at large radii ($R\sim 10^{11}~\textrm{cm}$) at the onset of Phase II, and attribute the structure of the X-ray light curve to the long-term evolution and slow central accretion of this extended disk.  We will see that the formation of the extended thin disk cannot be taken for granted due to the presence of a massive pressure-supported convective atmosphere around the inner disk. 

This work is organized as follows.  In Section \ref{sec:algorithm}, we discuss our numerical algorithm.  In Section \ref{sec:results}, we present the results our simulations.  In Section \ref{sec:discussion}, we present an analytical model for the neutrino-cooled central accretion that we do not resolve in the simulations, and provide a theory for the duration of the prompt accretion phase and the triggering of the steep decline of the X-ray light curve. We also attempt to extrapolate the evolution of the accretion rate beyond the duration of the simulations. Finally, in Section \ref{sec:conclusions}, we summarize our conclusions.

\section{Numerical Algorithm}
\label{sec:algorithm}

The simulations were carried out with the piecewise-parabolic solver in the adaptive-mesh-refinement code FLASH \citep{Fryxell:00}, version 2.5,  in two spatial dimensions using cylindrical coordinates $(R,z)$.  FLASH does not support angular momentum advection and viscous transport in this regime.  In Section \ref{sec:angular_momentum}, we describe our implementation of angular momentum transport.  In Section \ref{sec:initial_model}, we discuss our initial model and boundary conditions.  In Section \ref{sec:test}, we provide a test of angular momentum conservation.

\subsection{Angular Momentum Transport}
\label{sec:angular_momentum}

The specific angular momentum $\ell=R v_\phi$, where $v_\phi$ is the azimuthal velocity, was treated as a mass scalar quantity that was transported according to \citep[see, e.g.,][]{Pringle:81}
\bea
\label{eq:ell_evolve}
& &\frac{\partial (\rho \ell)}{\partial t} + \frac{1}{R}\frac{\partial( R v_R \rho \ell)}{\partial R}
+ \frac{\partial (v_z \rho \ell) }{\partial z} \nonumber\\ & &
\ \ \ \ \ \ \ \ \ \ \ \ \ \ \  - \frac{1}{R}\frac{\partial}{\partial R}\left[R^3 \nu \rho\frac{\partial}{\partial R}\left(\frac{\ell}{R^2}\right)\right] = 0 ,
\eea
where $\nu$ is a shear viscosity to be specified below.
Equation (\ref{eq:ell_evolve}), combined with the equation of continuity, is equivalent to the azimuthal axisymmetric Navier-Stokes equation
\beq
\frac{\partial v_\phi}{\partial t}+v_R\frac{\partial v_\phi}{\partial R} + \frac{v_R v_\phi}{R}+v_z \frac{\partial v_\phi}{\partial z} - \frac{1}{R^2\rho }\frac{\partial(R^2\nu\rho\sigma)}{\partial R} = 0 ,
\eeq
where 
\beq
\sigma = R \frac{\partial}{\partial R} \left(\frac{v_\phi}{R}\right)
\eeq
is the $R-\phi$ component of the shear tensor.  In FLASH, the calculation of the first three terms in equation (\ref{eq:ell_evolve})  is carried out through the mass scalar advection capability; the fourth, parabolic term is included explicitly in the calculation of the radial $\rho\ell$-flux in the code for the diffusion of mass scalars.

The energy dissipated through shear viscosity was accounted for by including the specific heating rate \citep[see, e.g.,][]{Landau:59}
\beq
\label{eq:viscous_heating}
\dot \epsilon_{\rm visc} = \nu \left[ R \frac{\partial}{\partial R} \left(\frac{\ell}{R^2}\right)\right]^2 = \nu \sigma^2 .
\eeq

Since we do not simulate the magnetic field of the fluid, we utilize a local definition of the shear viscosity to emulate the magnetic stress arising from the intrinsically nonlocal magnetorotational instability (MRI; \citealt{Balbus:98} and references therein).  It should be kept in mind, however, that the effects of MRI are in some respects very different from those of the viscous stress. For example, the thick disk surrounding our collapsar black hole is convective; in unmagnetized accretion flows convection transports angular momentum inward, toward the center of rotation \citep{Ryu:92,Stone:96,Igumenshchev:00b}, whereas in magnetized flows, convection can also transport angular momentum outward \citep{Balbus:02,Igumenshchev:02,Igumenshchev:03,Christodoulou:03}.  Thus our results must be interpreted with caution.  

Our definition of the local viscous stress emulating the MRI must be valid under rotationally supported, pressure supported, and freely falling conditions.  \citet{Thompson:05} suggest that since the wavenumber of the fastest growing MRI mode, which is given by the dispersion relation $v_{\rm A} k\sim \Omega$ where $v_{\rm A}$ is the Alfv\'en velocity and $\Omega=v_\phi/R$ is the angular velocity, should in the saturated quasi-state state be about the gas pressure scale height, $k\propto H^{-1}$, the Maxwell $\rho v_{\rm A}^2$ and viscous $\nu\rho \Omega$ stresses (up to factors in $|d\ln \Omega/d\ln R|$ that we neglect) can be equated if the viscosity is given by
\beq
\label{eq:viscosity_Thompson}
\nu_{\rm MRI} = \alpha H^2 \Omega ,
\eeq
where $\alpha$ is a dimensionless parameter. If the pressure scale height is defined locally, 
\beq
H=|\vec\nabla\ln P|^{-1} ,
\eeq
the viscosity defined in equation (\ref{eq:viscosity_Thompson}) suffers from divergences at pressure extrema.  To alleviate this problem, we define a second viscosity according to the \citet{Shakura:73} prescription 
\beq
\label{eq:viscosity_Shakura}
\nu_{\rm SS} = \alpha \frac{P}{\rho} \Omega^{-1} .
\eeq
Shakura-Sunyaev viscosity overestimates the magnetic stress in stratified hydrostatic atmospheres. We thus set the viscosity in equations (\ref{eq:ell_evolve}) and (\ref{eq:viscous_heating}) to equal the harmonic mean of the above two viscosities
\beq
\label{eq:harmonic_mean}
\nu = \frac{2~\nu_{\rm MRI}~\nu_{\rm SS} }{\nu_{\rm MRI}+\nu_{\rm SS}} .
\eeq
Our choice for the stress-to-total pressure ratio is $\alpha=0.01$, consistent with the ratio of the time-averaged stress to the time-averaged total pressure in the stratified, radiation-dominated disks in the simulations of \citet{Hirose:09}.  Hirose et al. found, however, that the fluctuations in the stress and the pressure (total or fluid) are not temporally coincident; this underscores the limitations of our assumed direct proportionality of the viscous stress with the total pressure.  In the limits $\nu_{\rm MRI}\gg \nu_{\rm SS}$ or $\nu_{\rm MRI}\ll \nu_{\rm SS}$, the effective value of the viscosity parameter implied by equation (\ref{eq:harmonic_mean})  is twice the nominal value, $\alpha_{\rm eff}\approx 0.02$.

Because FLASH employs an explicit method for the diffusion of mass scalars, numerical stability of the above viscous transport prescription places a stringent upper limit on the time step
\beq
\label{eq:tstep_diffuse}
\Delta t < \frac{\Delta R^2}{2\nu} ,
\eeq 
where $\Delta R$ is the grid resolution.  For $\alpha \gg 0.01$, the viscous time step in our simulations becomes prohibitively shorter than the Courant time step.  In our test simulations with a $\gamma$-law equation of state (EOS), we find that while not implying an outright instability, a choice of $\Delta t$ that saturates the limit in equation (\ref{eq:tstep_diffuse}) results in weak stationary staggered perturbations in the fluid variables.  We ignore this complication and allow our time step to be set by the limit in equation (\ref{eq:tstep_diffuse}) of the cell with the smallest viscous diffusion time across the cell.

The centrifugal force is included in the calculation of the gravitational acceleration via
\beq
\vec{a}_{\rm grav} = - \frac{G M_{\rm BH}}{r^3} \vec{r}  - \vec{\nabla} \Phi_{\rm fl} + \frac{\ell^2 }{R^4} \vec{R}
\eeq 
where $r=(R^2+z^2)^{1/2}$, $M_{\rm BH}$ is the mass of the central compact object which we take to be a black hole, and $\Phi_{\rm fl} (r)$ is the gravitational potential of the mass distribution of the fluid within the computational grid that has been spherically averaged around the origin $(R,z)=(0,0)$.  At the radii and densities that we resolve in the simulations, relativistic effects are weak; we thus treat the gravitational potential as Newtonian.  In Section \ref{sec:test} below, we present a test of the angular momentum transport code.

\subsection{Initial Model and Boundary Conditions}
\label{sec:initial_model}

The initial model is the rotating $\approx 14~M_\odot$ Wolf-Rayet star 16TI of \citet{Woosley:06a}, evolved to pre-core-collapse from a $16~M_\odot$ main sequence progentior.\footnote{\citet{LopezCamara:09} carried out SPH simulations of neutrino-cooled accretion during the first $0.5~\textrm{s}$ of the collapse and \citet{Morsony:07} simulated the propagation of a relativistic jet using the same model star.} To prepare the model 16TI, Woosley \& Heger assumed that the rapidly rotating progenitor, which is near breakup at its surface at $r\approx 4\times10^{10}~\textrm{cm}$, had low metallicity, $0.01~Z_\odot$, on the main sequence and became a WR star shortly after central H depletion, which implied an unusually small amount of mass loss.  For illustration, the specific angular momentum at the three-quarters mass radius was $\ell_{3/4}\sim 8\times10^{17} ~\textrm{cm}^2~\textrm{s}^{-1}$, implying circularization around a $10~M_\odot$ black hole at $R\sim 5\times10^8 ~\textrm{cm}$.  The circularization radii of the outermost layers of the star are in the range $10^9-10^{10} ~\textrm{cm}$.  Woosley \& Heger provide a radius-dependent angular momentum profile $\ell(r)$; we endowed this with a dependence on the polar angle $\theta\equiv \cos^{-1} (z/r)$ via 
\beq
\ell(r,\theta) = \ell(r) \sin^2 (\theta) , 
\eeq
such that the star rotates rigidly on spherical shells.

We placed the center of the star at the origin, $(R,z)=(0,0)$.  
Pseudo-logarithmic gridding was achieved by capping the adaptive resolution at radius $r$ with
$\Delta R, \ \Delta z > \frac{1}{8} \eta r$, where we choose $\eta=0.05$; this
prevents use of excessive resolution far from the center of the star.  
Beyond the outer edge of the star at $r_{\rm star}=4\times10^{10} ~\textrm{cm}$ we placed a cold ($10^4~\textrm{K}$) low-density stellar-wind like medium with density profile $\rho(r)=3\times10^{-7} ~ (r/r_{\rm star})^{-2}~\textrm{g cm}^{-3}$.  Since the model 16TI of \citet{Woosley:06a} was not constructed to be in hydrostatic equilibrium in the presence or rotation, we ignored rotation and set the initial density distribution to be spherically symmetric at the beginning of the simulations.  This is a poor approximation in the very outer layers of the star, as is evident from the ellipsoidal distortion that sets in at the beginning of the simulation.

For the equation of state we chose the Helmholtz EOS provided with the FLASH distribution \citep{Timmes:00}, which contains contributions to pressure and internal energy from radiation, ions, electrons, positrons, and Coulomb effects.  We passively advected the abundances of seven nuclear species represented in the model including $^4$He, $^{12}$C, $^{16}$O, $^{20}$Ne, $^{24}$Mg, $^{28}$Si, and $^{56}$Fe.  The local nuclear composition was passed to the Helmholtz EOS as input.  We do not simulate nuclear reactions, nuclear photodisintegration, and neutrino emission and absorption.  These processes are certainly important in the hot inner accretion disk around the collapsar black hole, but since we simulate only the outer, cooler disk with temperatures $T< 10^{10}~\textrm{K}$, the neglect of nuclear and neutrino processes is a reasonable approximation.

\begin{deluxetable*}{crrrrrrrrr}
\tablecolumns{10}
\tablewidth{7in}
\tablecaption{Summary of Simulation Parameters and Key Measurements\label{tab:simulations}}
\tablehead{
  \colhead{Run Number} & 
  \colhead{$R_{\rm min}\ (\textrm{cm})$ \tablenotemark{a}} & 
  \colhead{$(R,z)_{\rm max}\ (\textrm{cm})$ \tablenotemark{b}} & 
  \colhead{$\Delta (R,z)_{\rm min}\ (\textrm{cm})$ \tablenotemark{c}} & 
  \colhead{$M_{\rm BH,init}\ (M_\odot)$ \tablenotemark{d}} & 
  \colhead{$t_{\rm max}\ (\textrm{s})$ \tablenotemark{e}} &
  \colhead{$\alpha$ \tablenotemark{f}} &
  \colhead{$t_{\rm decl}\ (\textrm{s})$ \tablenotemark{g}} &
  \colhead{$M_{\rm BH,decl}\ (M_\odot)$ \tablenotemark{h}} &
  \colhead{$(d\ln \dot M/d\ln t)_{\rm decl}$ \tablenotemark{i}}
}
\startdata
1 & $5\times10^7$ & $5\times10^{11}$ & $1.9\times10^6$ & $0.51$ & $10^2$ & $0.01$ & $37$ & $7.35$ & $-2.8$ \\
2 & $10^8$ & $10^{11}$ & $6.1\times10^6$ & $1.26$ & $10^3$ & $0.01$ & $47$ & $8.97$ & $-2.7$ \tablenotemark{j} \\
3 & $2\times 10^8$ & $5\times10^{11}$ & $7.6\times10^6$ & $2.05$ & $2\times10^3$ & $0.01$ & $52$ & $10.44$ & $-2.3$ \tablenotemark{k} 
\enddata
\tablenotetext{a}{The minimum cylindrical radius.}
\tablenotetext{a}{The maximum cylindrical radius and absolute vertical latitude.}
\tablenotetext{c}{The minimum resolution element size.}
\tablenotetext{d}{Initial black hole mass.}
\tablenotetext{e}{Duration of the simulation.}
\tablenotetext{f}{The viscous stress parameter (see Section \ref{sec:angular_momentum}).}
\tablenotetext{g}{Time of the beginning of the steep decline of the accretion rate.}
\tablenotetext{h}{Black hole mass at the beginning of the steep decline of the accretion rate.}
\tablenotetext{i}{Logarithmic slope of the decline of the accretion rate.}
\tablenotetext{j}{For $50~\textrm{s}\leq t\leq 500~\textrm{s}$.}
\tablenotetext{k}{For $52~\textrm{s}\leq t\leq 200~\textrm{s}$.}
\end{deluxetable*}

The simulation was carried out in the annular cylindrical domain $R_{\rm min}< R<R_{\rm max}$ and $-z_{\max} < z < z_{\rm max}$.  Because the impact of time-step limitations (eq.\ [\ref{eq:tstep_diffuse}]) on the computational cost, and to avoid dealing with the fluid hot enough to be susceptible to efficient neutrino cooling, the smallest inner radius $R_{\rm min}$ that we could simulate was $R_{\rm min}\sim 5\times10^7~\textrm{cm}$.  We placed the outer boundaries well outside the star $R_{\rm max}=z_{\rm max}=(1-5)\times10^{11}~\textrm{cm}$.  In Table \ref{tab:simulations}, we summarize the main parameters of our simulations, and also present some of the key measurements, defined in Section \ref{sec:results}, characterizing the outcome of each simulation.  Each simulation was run for $\sim 10^6$ hydrodynamic time steps and consumed $\sim 20,000$ CPU hours on the Texas Advanced Computing Center's clusters Lonestar and Ranger.

The boundary condition at $R_{\rm min}$ was unidirectional outflow that allowed free flow from larger to smaller radii (off the grid) and disallowed flow from smaller to larger radii (onto the grid) by imposing a reflecting boundary condition at $R_{\rm min}$ whenever $v_R$ was positive in the leftmost grid cell.    We imposed the torque-free boundary condition\footnote{A motivation of the torque-free boundary condition can be found in \citet{Zimmerman:05}.} via 
\beq
\label{eq:torque_free}
\frac{\partial}{\partial R}\left(\frac{\ell}{R^2}\right)_{R=R_{\rm min}} = 0 .
\eeq
As in other Eulerian codes, the boundary conditions in FLASH are imposed by assigning values to fluid variables in rows of ``guard'' cells just outside the boundary of the simulated domain.  At any given value of $z$ on the computational grid, let $R_{1/2}$ denote the leftmost cell within the simulated domain, and let $R_g$ where $g=(-\frac{7}{2},-\frac{5}{2},-\frac{3}{2},-\frac{1}{2})$ be the four guard cells to the left of $R_{1/2}$ such that the grid separation corresponds to $\Delta g=1$. The torque-free boundary condition, if assumed to apply for $R\leq R_{\rm min}$, implies $\ell_g/R_g^2=\ell_{1/2}/R_{1/2}^2$.  We fixed the guard-cell velocity perpendicular to the left vertical boundary to $v_{R,g} = -|v_{R,1/2}| R_{1/2}/ R_g$, which, with the assumption of uniform density $\rho_g=\rho_{1/2}$, ensures mass continuity in the guard cells and the vanishing of the mass flux across $R=R_{\rm min}$ if $v_{R,1/2}>0$.  All other fluid variables $X$ were simply copied into the guard cells, $X_g=X_{1/2}$, and were subsequently rendered thermodynamically consistent.  This simple prescription \emph{approximates} free inflow (toward smaller $R$) across $R_{\rm min}$, but of course, the guard cell values violate energy and momentum conservation at $R<R_{\rm min}$.

The mass of the black hole $M_{\rm BH}$ was initialized with the stellar mass initially located outside the grid, at $R<R_{\rm min}$.  The black hole mass was evolved by summing mass flux crossing the boundary at $R=R_{\rm min}$, 
\beq 
\frac{dM_{\rm BH}}{dt}=-2\pi R_{\rm min}\int_{-z_{\rm max}}^{z_{\rm max}} v_R(R_{\rm min},z) \rho(R_{\rm min},z) dz .
\eeq

\subsection{Test of the Code}
\label{sec:test}

\begin{figure}[t]
\begin{center}
\includegraphics[width=3.5in]{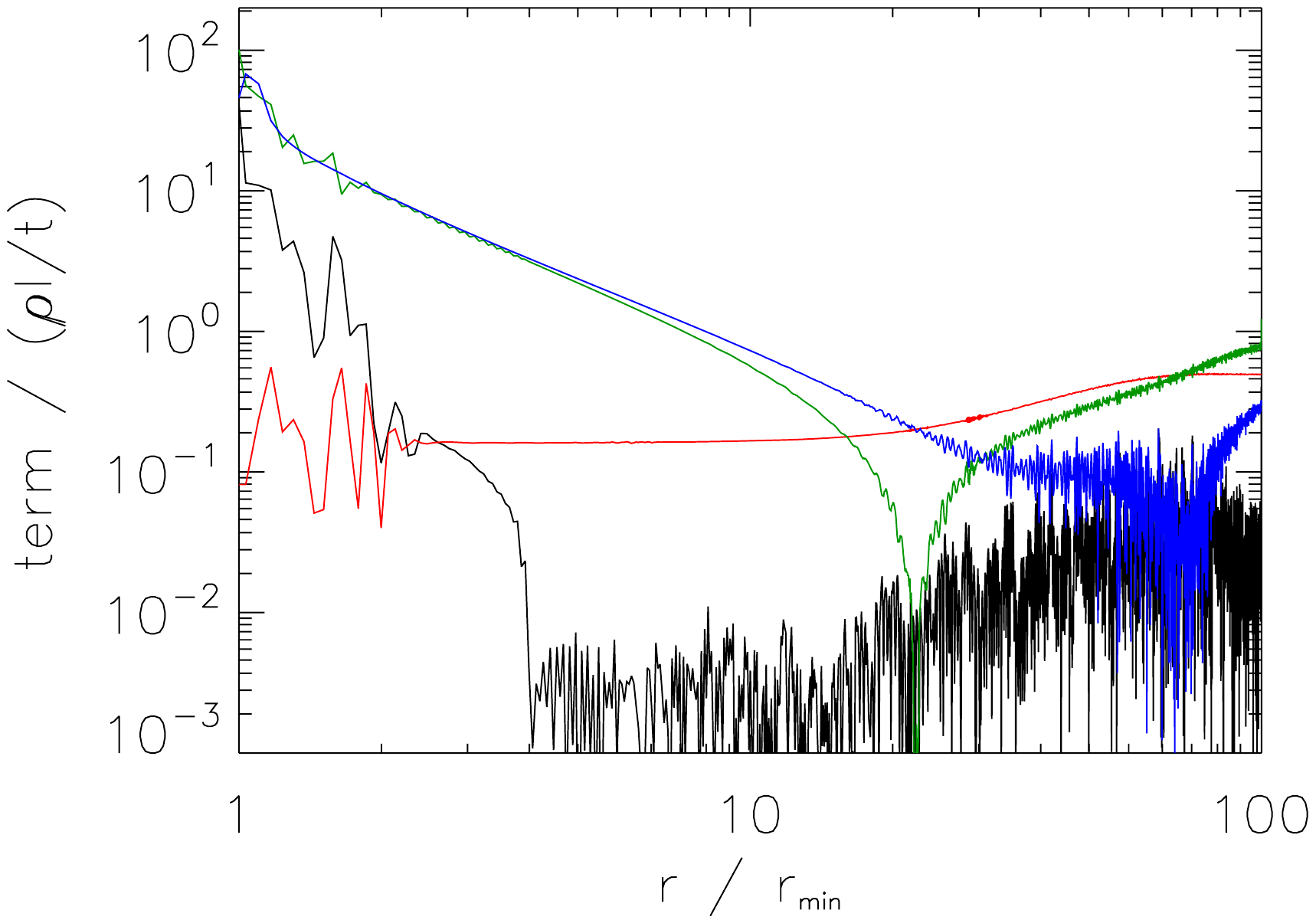}
\end{center}
\caption{A test of angular momentum transport, showing the magnitude of the Eulerian time derivative ($\tau_1$, {\it red curve}), advection ($\tau_2$, {\it green curve}), and viscous transport ($\tau_3$, {\it blue curve}) terms in the one-dimensional ($\partial/\partial z=0$) version of the angular momentum conservation relation in equation (\ref{eq:ell_evolve}).  The terms are expressed in the units of $\rho\ell/t$.  We also show the sum $|\tau_1+\tau_2+\tau_3|$ ({\it black curve}), which should be much smaller than the largest of the three terms.  The violation of angular momentum conservation near the left radial boundary is associated with the conservation-violating nature of the torque-free boundary condition that we have imposed.}
\label{fig:test}
\end{figure}

To test our implementation of angular momentum transport, we performed a one-dimensional ($\partial/\partial z=0$) simulation of an initially uniform temperature and density fluid with the $\gamma=\frac{5}{3}$ equation of state orbiting in a Keplerian potential.  We use a uniform radial grid with $R_{\rm max}=100\ R_{\rm min}$ and grid spacing $\Delta R=0.06~R_{\rm min}$.  The initial temperature was chosen such that the sound speed was about $16\%$ of the Keplerian velocity at the inner radial boundary, and $1.6$ times the Keplerian velocity at the outer radial boundary.  The time step was limited by $\Delta t\leq \frac{1}{4}\Delta R^2/\nu$, which is a factor of two more stringent than the stability condition in equation (\ref{eq:tstep_diffuse}). We found that reducing the time step to a half of the value required for stability substantially reduces, but does not entirely eliminate, the noise in the error estimator that we are about to discuss. We evaluate the nonzero terms in equation (\ref{eq:ell_evolve}) directly from the numerical data.  Let the $\tau_1$, $\tau_2$, and $\tau_3$ denote the first, second, and fourth term in equation (\ref{eq:ell_evolve})
\bea
\tau_1&\equiv&\frac{\partial (\rho \ell)}{\partial t}, \nonumber\\
\tau_2&\equiv& \frac{1}{R}\frac{\partial( R v_R \rho \ell)}{\partial R} ,\nonumber\\
\tau_3&\equiv& - \frac{1}{R}\frac{\partial}{\partial R}\left[R^3 \nu \rho\frac{\partial}{\partial R}\left(\frac{\ell}{R^2}\right)\right] .
\eea
Correct angular momentum transport requires 
\beq
\label{eq:test_condition}
|\tau_1+\tau_2+\tau_3|\ll {\rm max} (|\tau_1|,|\tau_2|,|\tau_3|) .
\eeq

In Figure \ref{fig:test}, we plot $|\tau_1|$, $|\tau_2|$, $|\tau_3|$, and $|\tau_1+\tau_2+\tau_3|$ over the entire range of radii after $\sim 1,000$ Keplerian orbital periods at $R_{\rm min}$, which corresponds to $\sim 1$ orbital period at $R_{\rm max}$.  Radial derivatives were computed by 3-point Lagrangian interpolation with the IDL routine {\tt DERIV}.  In computing the derivative at inner boundary, we included guard cells in manner equivalent with the boundary condition prescription used in our 2D simulations, as described in Section \ref{sec:initial_model} above.  Some violation of the transport equation (\ref{eq:ell_evolve}) is expected at the two leftmost grid cells at $R\approx (1.03,1.09)\times R_{\rm min}$ because the torque-free boundary condition does not conserve angular momentum.  Apart from the leftmost cells, the angular momentum is conserved at the $10\%$ level or better at all radii.  The spatial derivative in the viscous transport term ($\tau_3$) is partially responsible for the noise evident at large radii.  The noise at small radii seems to be correlated with the viscous time step limiter, which suggests that it is related to the explicit nature of our viscous diffusion scheme.

\section{Results}
\label{sec:results}

\begin{figure}
\begin{center}
\includegraphics[width=0.45\textwidth,clip]{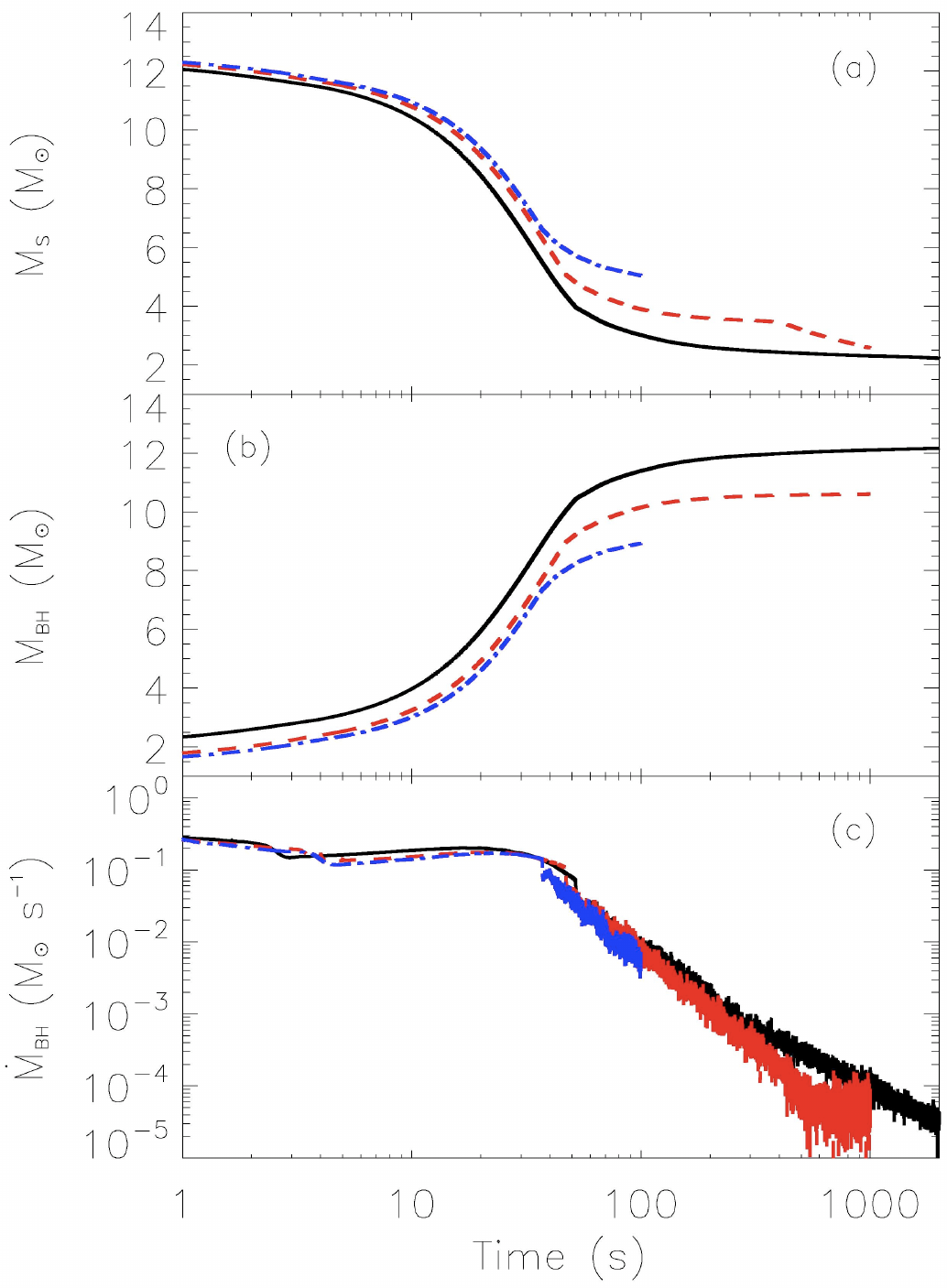}
\end{center}
\caption{(a) Stellar mass that remains in the simulation as a function of time in Run 1 (\emph{blue, dot-dashed line}), Run 2 (\emph{red, dashed-line}), and Run 3 (\emph{black, solid-line}).  The drop at $t\approx400\textrm{s}$ in Run 2 is an artifact of fluid escape from the box through the boundaries at $R=R_{\rm max}$ and $z=\pm z_{\rm max}$. (b) Mass of the black hole.  (c) The rate at which fluid mass accretes across the boundary at $R=R_{\rm min}$ and is added to the mass of the black hole.  The sharp drop at $t\approx37-52~\textrm{s}$ coincides with the formation of the accretion shock and the onset of convection in the rotationally and hydrostatically supported fluid.  The flattening at $t\sim 500~\textrm{s}$ in Run 2 coincides with the cessation of the accretion of low angular momentum fluid through the axial funnel.  The power-law accretion rate decline in Run 3 exhibits a shallowing of the logarithmic slope at $t\sim 200~\textrm{s}$.}
\label{fig:mdot}
\end{figure}

Since the evolution of the central accretion rate seems to fall into three distinct phases (see Figure \ref{fig:mdot})  that appear to correspond to the phases identified in the GRB $\gamma$-ray and X-ray light curve \citep[e.g.,][]{ZhangB:06}, we divide our description of the results of the simulations into three parts.  In Section \ref{sec:phase_0}, we describe Phase 0 that concludes with the appearance of an accretion shock.  We also discuss the limitations of our method in this regime, and set the stage for an analytic model that we develop further below in Section \ref{sec:triggering_decline} to take into account the physics left out the our simulations.  In Section \ref{sec:phase_1}, we describe Phase I that is characterized by a steep, power-law decline of the central accretion rate and a rapid hydrodynamic readjustment of the accreting stellar envelope.  In Section \ref{sec:phase_2}, we describe Phase II, in which the central accretion rate steadies.  The corresponding ``plateau'' phase in GRB X-ray light curves eventually ends and gives way to a renewed steeper decline.  Because of computational cost limitations, we do not extend our runs to $\sim10^4~\textrm{s}$, where, based on the observed light curves, one would expect the renewed steeper decline to occur, but further below, in Section \ref{sec:long_term}, we briefly speculate about the long-term evolution of the light curve.

\subsection{Phase 0: Quasiradial Accretion}
\label{sec:phase_0}

\begin{figure*}
\begin{center}
\includegraphics[width=0.45\textwidth,clip]{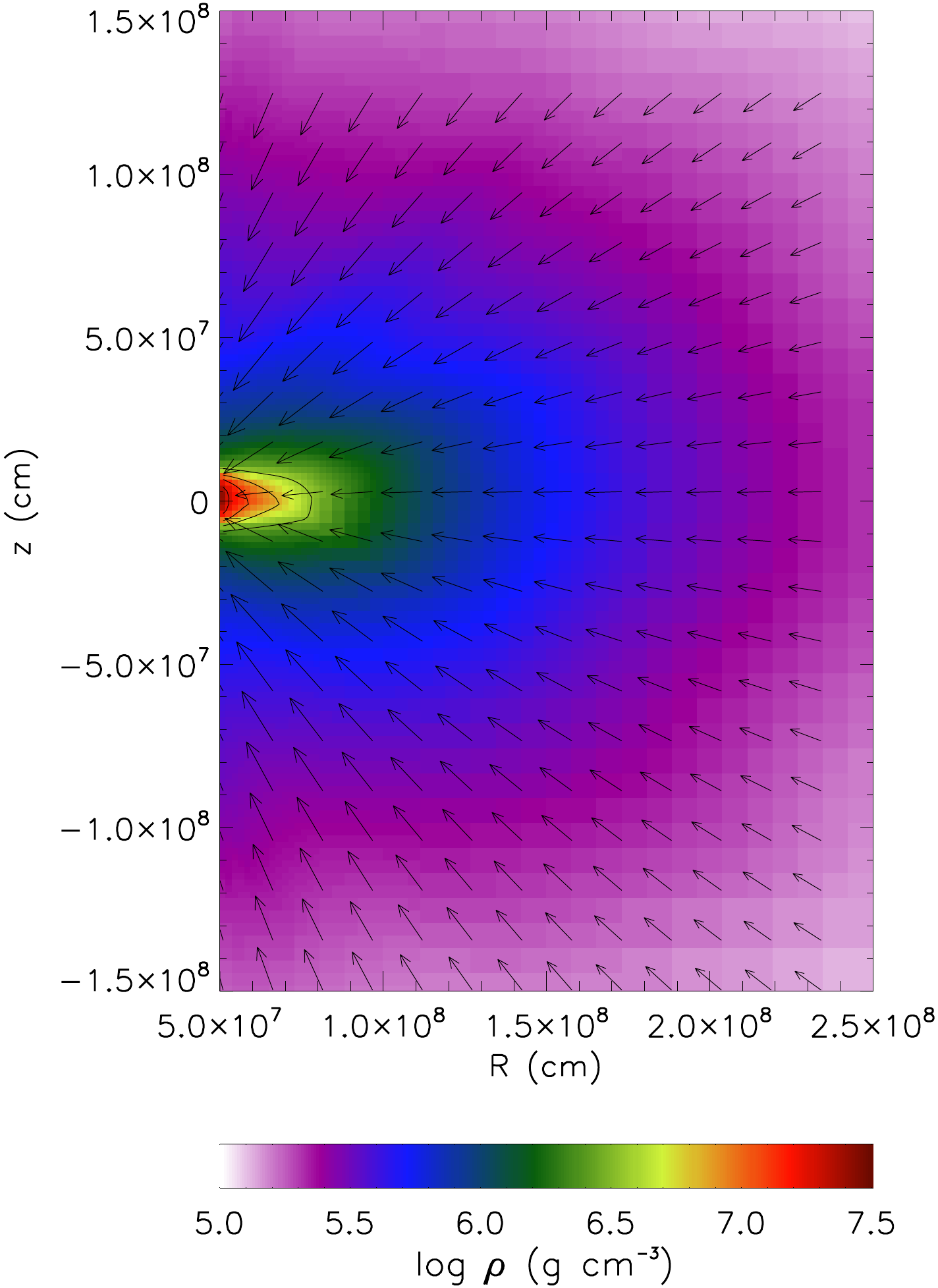}
\includegraphics[width=0.45\textwidth,clip]{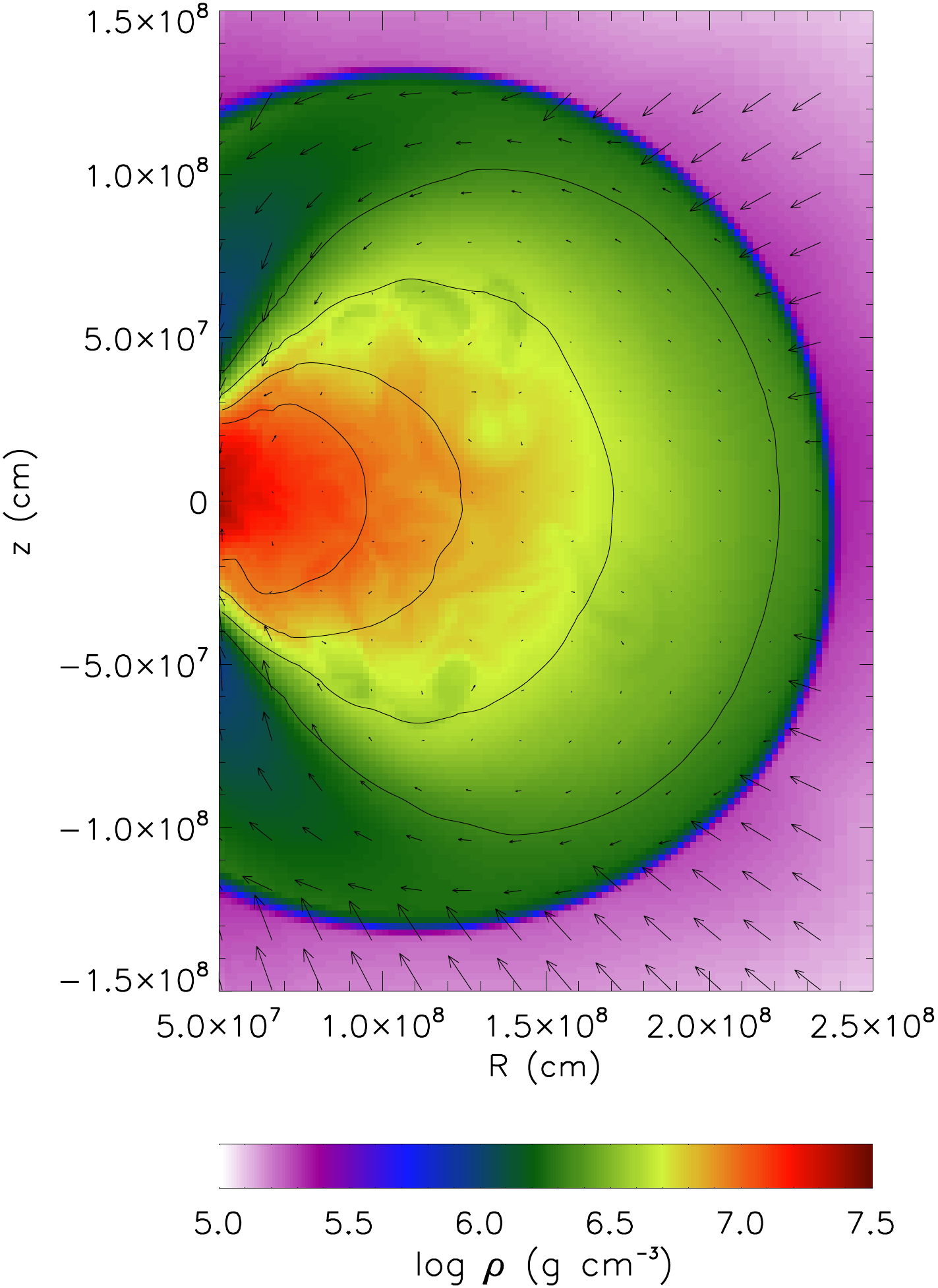}
\end{center}
\caption{The innermost accretion flow, composed mostly of oxygen and neon, shortly prior (\emph{left panel}) and shortly following (\emph{right panel}) the formation of an accretion shock at $t\approx 37~\textrm{s}$ in Run 1 with $R_{\rm min}=5\times10^7~\textrm{cm}$.  The color scale denotes the fluid density, which in this region ranges between $10^5-10^{7.5}~\textrm{g cm}^{-3}$.  The black contours show the $T=(4,5,6,7)\times10^9~\textrm{K}$ isotemperature contours.  The arrows show the meridional component of the fluid velocity; the longest arrows correspond to $(v_R^2+v_z^2)^{1/2}=5\times10^9~\textrm{cm s}^{-1}$ and Mach numbers $\sim10$. Following shock formation, supersonic inflow resumes along the axial funnel.}
\label{fig:shock}
\end{figure*}

At the radii $R>R_{\rm min}$ that we resolve in the simulations, the stellar collapse proceeds quasiradially for a number of seconds until an accretion shock appears at the innermost simulated point at $(R,z)=(R_{\rm min},0)$.  The appearance of an accretion shock coincides with the emergence of a rotationally supported flow in the zones at the smallest radii.  In Figure \ref{fig:shock}, left panel, we show the density distribution, temperature contours, and velocity field during the quasiradial accretion phase, at $t=37~\textrm{s}$, just prior to the formation of the accretion shock, in the simulation with $R_{\rm min}=5\times10^7~\textrm{cm}$ (Run 1, see Table \ref{tab:simulations}).   The maximum density and temperature in this snapshot are $2.6\times10^7~\textrm{g cm}^{-3}$ and $T=7.5\times10^9~\textrm{K}$.  Since the simulation does not allow for nuclear photodisintegration, the temperature in the innermost cells in the simulation is an overestimate.  

The existence of a quasiradial accretion phase and the late formation of an accretion shock are clearly artifacts of the choice not to simulate the innermost $5\times10^7~\textrm{cm}$ from the central axis.  This innermost resolved radius is still $\sim100$ times larger than the gravitational radius of the nascent black hole.  Simulations that resolved the innermost radii at or near the ISCO \citep{MacFadyen:99,Proga:03c,Lee:06,Nagataki:07,Barkov:08,Komissarov:09,LopezCamara:09}, but were run much shorter than ours, saw accretion shock formation much earlier, during the first second of the collapse.  Some of the material falling quasiradially during the initial phase has enough angular momentum to circularize at radii that we do not resolve, but that are still larger than the ISCO. Indeed, our accretion shock forms earlier in runs with smaller $R_{\rm min}$ (see Table \ref{tab:simulations} and Figure \ref{fig:mdot}), consistent with the observation that circularization triggers shock formation.   Therefore, in general, the accretion shock forms when the orbital pericenter of the material crossing the equatorial plane becomes larger than the ISCO, or the innermost resolved radius $R_{\rm min}$ in the simulations in which the ISCO is unresolved. 

In Section \ref{sec:triggering_decline}, we will analyze differences between the inner accretion flow in our adiabatic simulations and that in the realistic GRBs progenitors, and suggest that the steep decline of the accretion rate in realistic GRB progenitors is triggered by the onset of circularization of the infalling stellar material at radii where the post-accretion-shock temperature is too low to allow for efficient cooling by neutrino emission.  We will conclude that the decline seems to be associated with the onset of outward expansion of the accretion shock.  The outward expansion is distinct from and could occur much later than the first occurrence of the shock.  In Section \ref{sec:triggering_decline} we will present a crude analytic model in which we estimate that the triggering of the steep decline should occur at $t_{\rm decl}\sim20~\textrm{s}$ in stars with density and angular momentum stratification as in 16TI.  This estimate is somewhat shorter than the shortest interval $t_{\rm decl}\sim 37~\textrm{s}$ observed in the highest-resolution simulation, Run 1.

\begin{figure*}[h]
\begin{center}
\includegraphics[width=0.45\textwidth,clip]{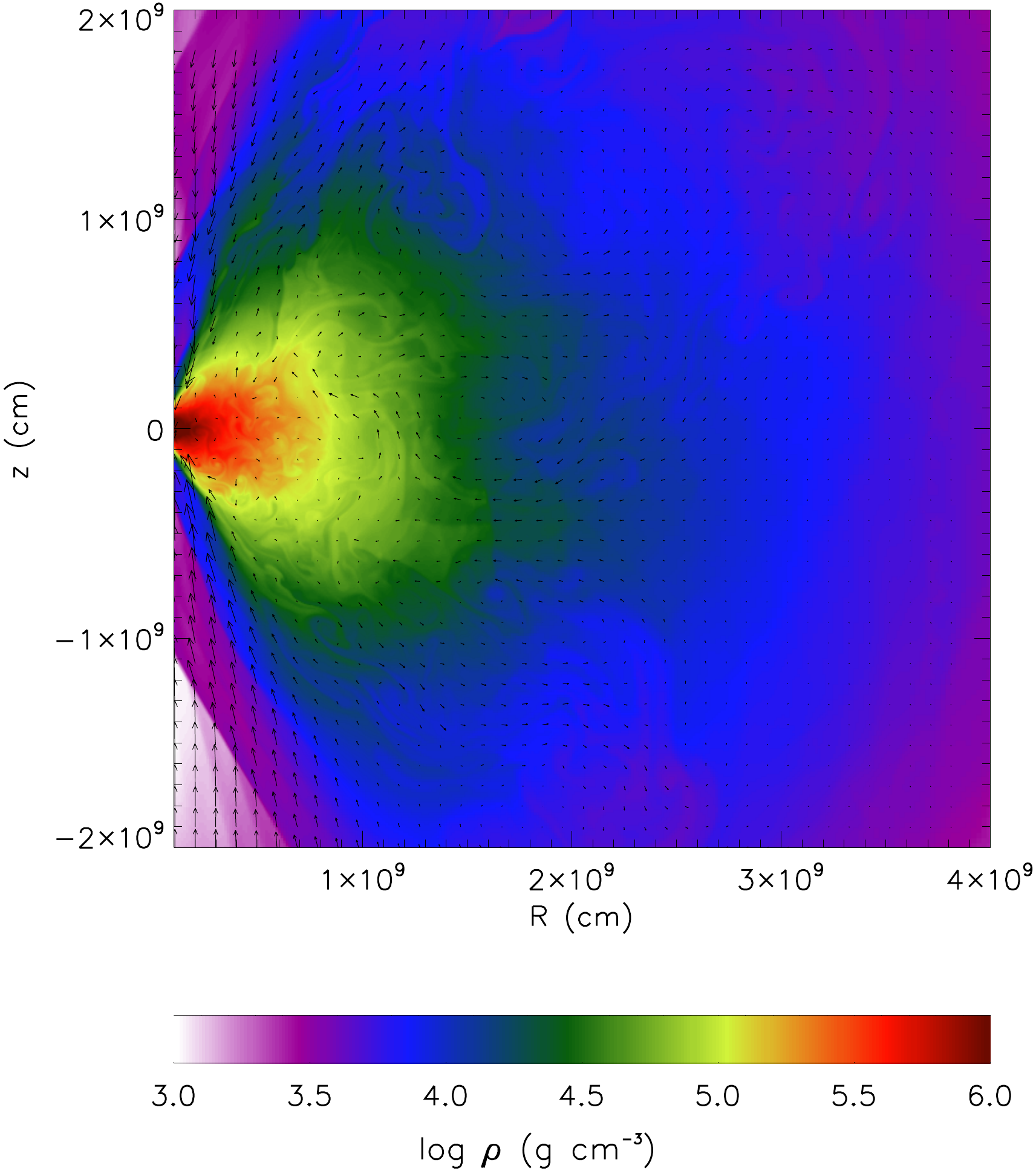}
\includegraphics[width=0.45\textwidth,clip]{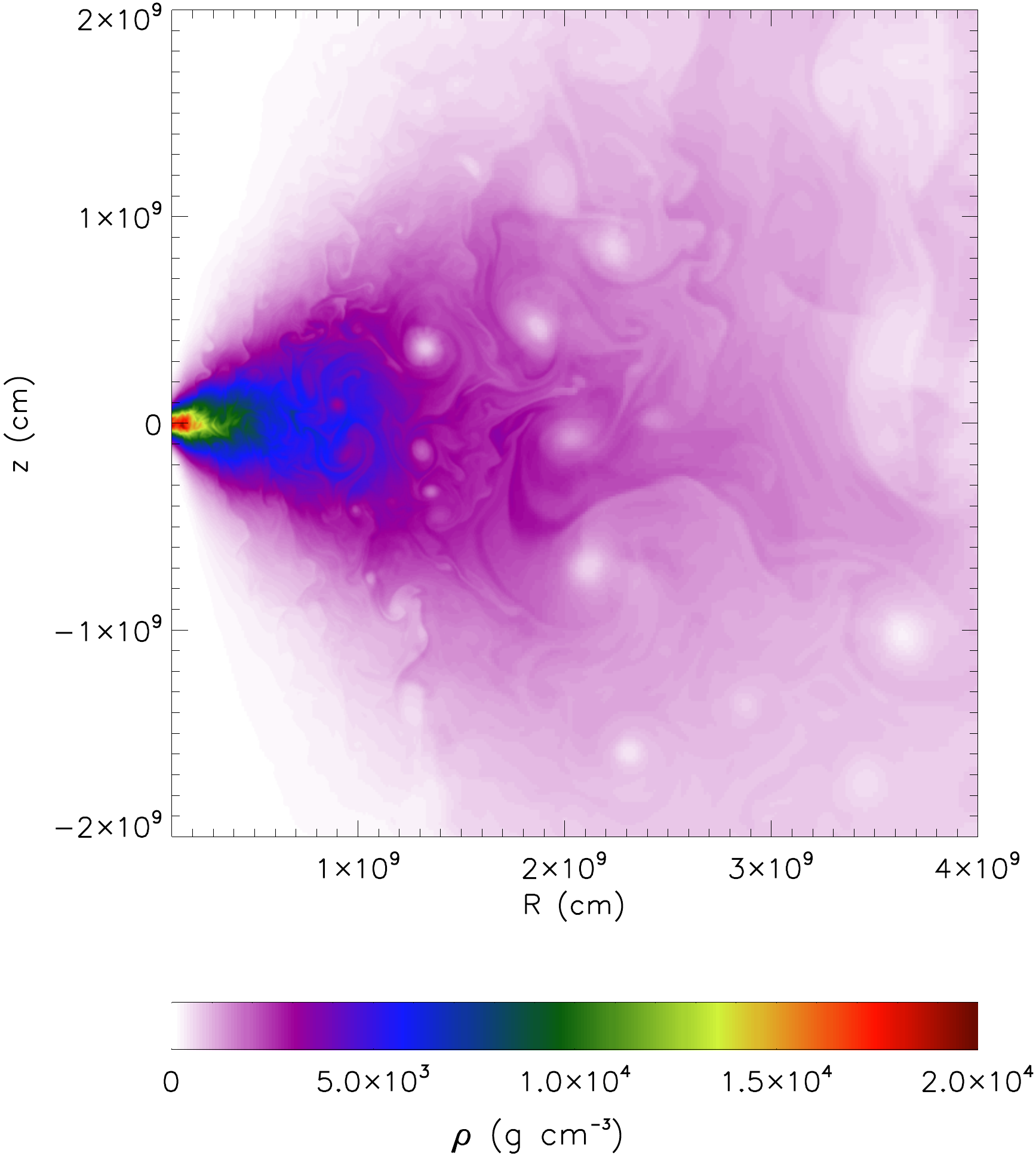}
\end{center}
\caption{The fluid density at $t=100~\textrm{s}$ (\emph{left panel}, logarithmic rendering) and $t=1,000~\textrm{s}$ (\emph{right panel}, linear rendering) in Run 2.  At early times, the fluid accreting supersonically through the axial funnel traverses multiple weak standing accretion shocks before it joins the disk or passes the boundary at $R=R_{\rm min}$.}
\label{fig:density}
\end{figure*}

\begin{figure*}[h]
\begin{center}
\includegraphics[width=0.45\textwidth,clip]{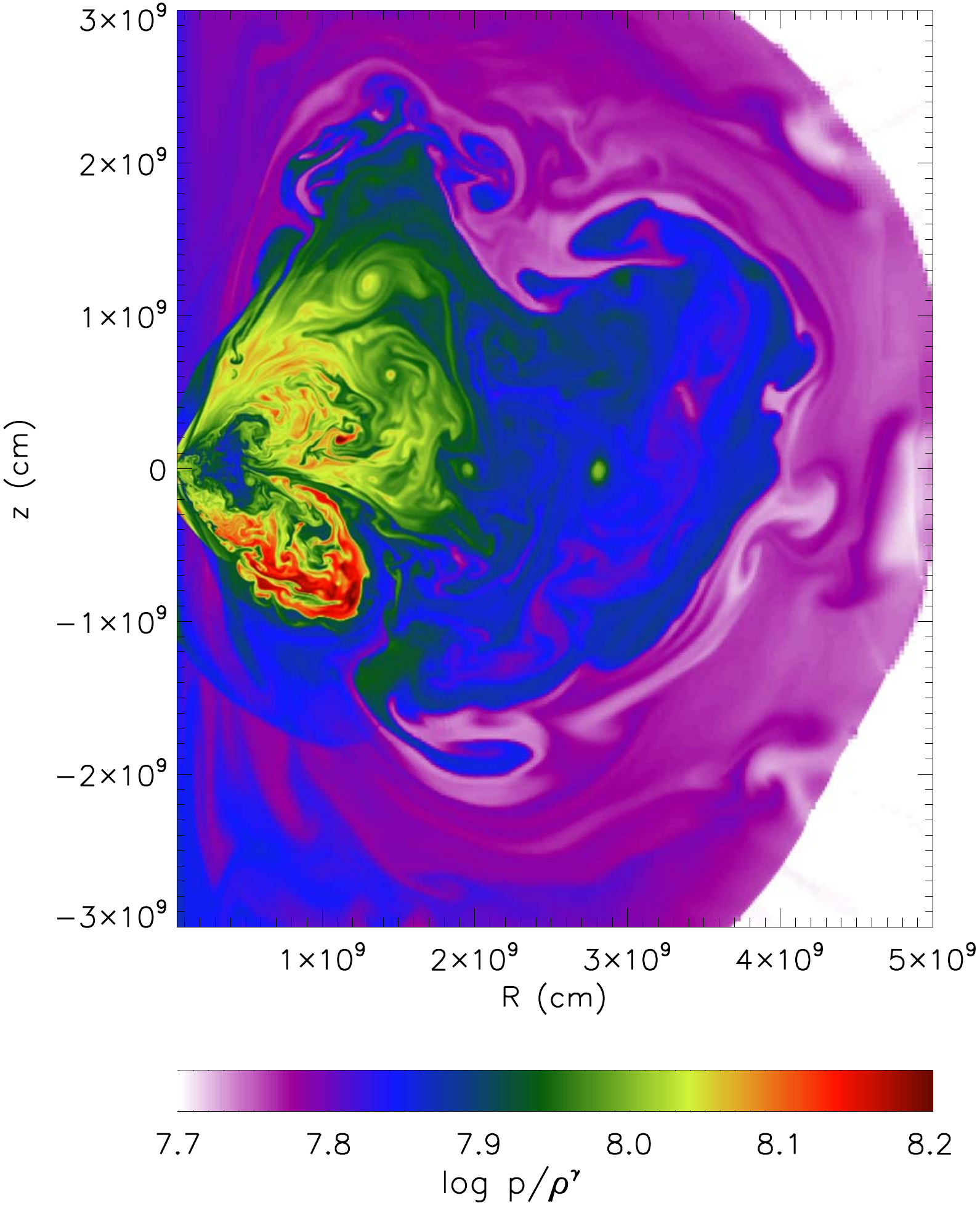}
\includegraphics[width=0.45\textwidth,clip]{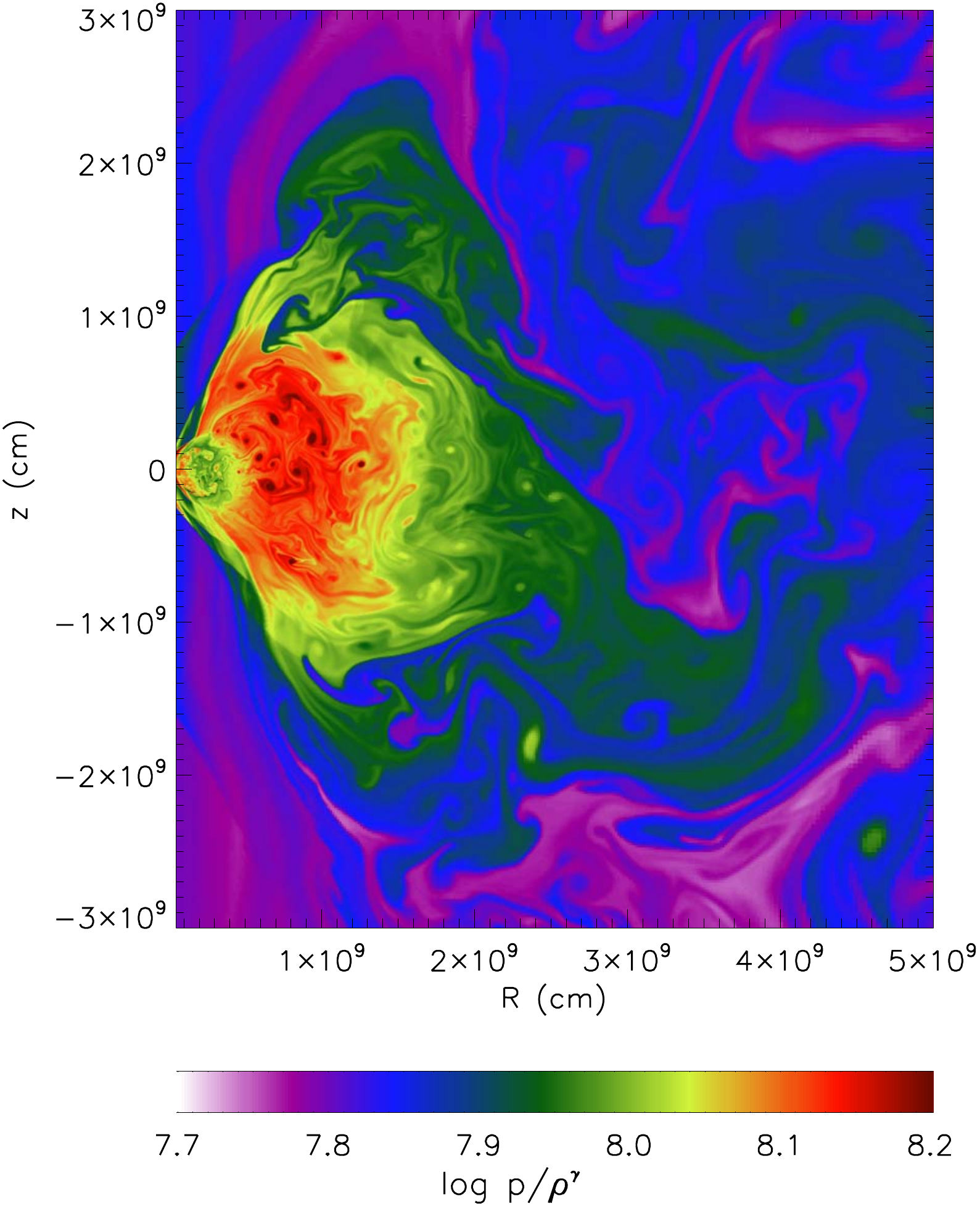}
\end{center}
\caption{The quantity $p/\rho^\gamma$ which is related to the specific entropy of the fluid, where $\gamma\equiv (d\ln p/d\ln \rho)_{s={\rm const}}$ at constant entropy, of the fluid at $t=70~\textrm{s}$ (\emph{left panel}) and $t=100~\textrm{s}$ (\emph{right panel})  in the center of the star, in Run 1 with $R_{\rm min}=5\times10^7~\textrm{cm}$.  The high entropy fluid tracks the outflow from the disk.  The low-entropy fluid accreting through the axial funnel traverses multiple weak standing accretion shocks before it joints the disk or passes the boundary at $R=R_{\rm min}$.  The primary, outward propagating accretion shock is visible along the right edge of the left panel.}
\label{fig:entropy}
\end{figure*}

\begin{figure}[t]
\begin{center}
\includegraphics[width=0.45\textwidth,clip]{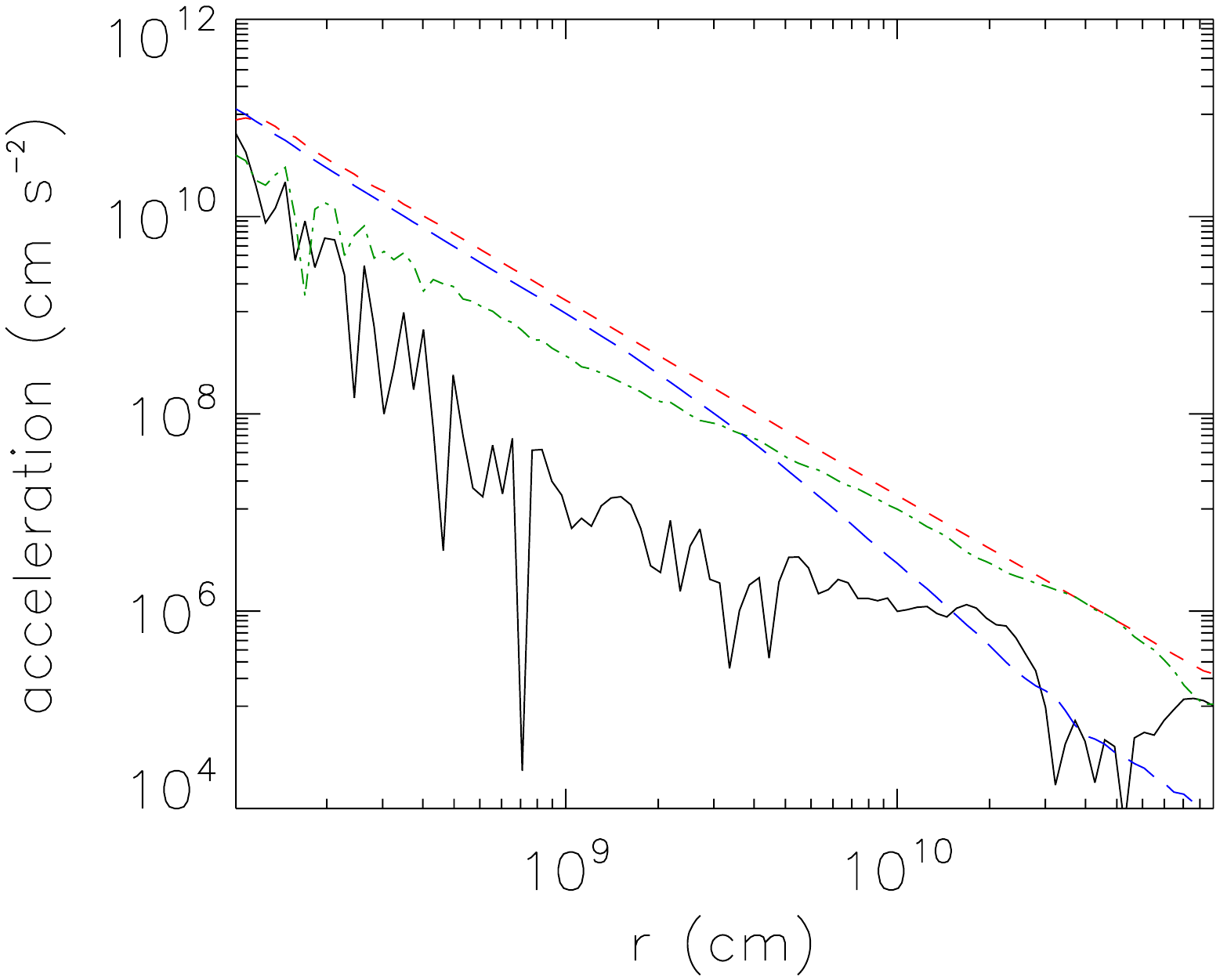}
\end{center}
\caption{Gravitational field $-\partial \Phi/\partial r$ (\emph{red short dashed line}), the pressure acceleration $-\rho^{-1} \partial p/\partial r$ (\emph{green dot-dashed line}), centrifugal acceleration in the radial direction $v_\phi^2 {\hat R}\cdot {\hat r}/r$ (\emph{blue long-dashed line}), and the sum of the gravitational, pressure, and centrifugal acceleration (\emph{black solid line}) in Run 2 with $R_{\rm min}=10^8~\textrm{cm}$.  The flow is rotationally supported at $r\lesssim 3\times10^9~\textrm{cm}$ and pressure supported at $r\gtrsim 4\times10^9~\textrm{cm}$.  We utilized the averages of $p$ and $\rho$ on spherical shells of radius $r$, and the mass-weighted averages of $\Phi$ and $v_\phi$ on spherical shells.  Furthermore, time-averaging was carried out in the interval $t=600-1,000~\textrm{s}$.}
\label{fig:acceleration}
\end{figure}

\subsection{Phase I: Funnel and Thick Disk Accretion}
\label{sec:phase_1}

At $t\sim 37-52~\textrm{s}$ where the shortest time scale corresponds to the simulations that resolve the smallest radii, an accretion shock forms along the equator near the inner boundary (Figure \ref{fig:shock}) and travels outward with a velocity $\sim 5\times10^7~\textrm{cm s}^{-1}$.  The shocked fluid at polar angles $|\theta-\pi/2|\lesssim 75^\circ$ is rotationally supported.   Figure \ref{fig:density}, left panel, shows that the isodensity contours of this rotationally supported fluid are roughly circular; the vertical and the cylindrically radial pressure scale heights are comparable.   The shocked fluid is turbulent and apparently convective (in two spatial dimensions, long-lived vortices form in the shock; the persistence of the vortices is an artifact of the assumed axisymmetry).  The maximum temperature in the post shock fluid is $9.3\times10^9~\textrm{K}$; this is a temperature at which the nuclear and neutrino physics that we ignore is marginally important; our neglect of photodisintegration cooling implies an overestimate of the temperature in the inner thick disk, $r\lesssim 10^8~\textrm{cm}$.  The shocked fluid in the $|\theta -\pi/2| \gtrsim 75^\circ$ cone around the vertical axis continues to infall supersonically.   

Figure \ref{fig:entropy} shows the quantity $p/\rho^\gamma$, which is related to the specific entropy, where $\gamma\equiv (d\ln p/d\ln \rho)_{s={\rm const}}$ at $t=70~\textrm{s}$ and $t=100~\textrm{s}$ in the highest resolution run, Run 1.  Entropy appears to be generated throughout the thick disk.  The high-entropy fluid exhibits a flow morphology suggestive of a ``disk wind.''  The strongest outflow tracked by the highest entropy fluid is along the interface of the turbulent thick equatorial disk and the supersonic axial inflow.  Prior to the cessation of the axial inflow, the wind streamlines do not terminate at infinity, but rather bend back toward the equatorial plane, suggesting a closed meridional circulation pattern that transports the energy generated in the thick disk.  \citet{Ohsuga:05} and \citet{Lee:06} have previously observed such a large-scale circulation pattern in their simulations.  The high entropy fluid appears to accumulate at the interface of the thick, rotationally-supported disk and the pressure-supported atmosphere, and to mix convectively in the atmosphere.

The appearance of the accretion shock is accompanied by a sudden rapid power-law decline of the central accretion rate.  The times of shock formation and the onset of decline in different simulations are provided in Table \ref{tab:simulations}.  Evolution of the black hole mass, the residual stellar fluid mass, and the accretion rate, is shown in Figure \ref{fig:mdot}.  The steep decline of the accretion rate resembles the rapid decline ubiquitous in the observed GRB X-ray light curves.  In the simulations, the decline starts at $\sim37-52~\textrm{s}$ and lasts until $\sim 200-500~\textrm{s}$.  The logarithmic derivative of the central accretion rate during the decline is $d\ln \dot M/d\ln t \approx-2.3-(-2.8)$, with the steepest decline corresponding to Run 1, the simulation that resolves the smallest radii.  

There does not seem to be a single explanation for the steepness of the decline of the accretion rate.  We have been able to identify three processes that seem to contribute.  We focus on Run 2, the run with the highest central resolution that we have run long enough to witness the end of the decline.

First, the Eulerian density within the thick disk decreases by a factor of $\sim 50-100$ from $t= 50~\textrm{s}$ to $500~\textrm{s}$.  The density drop occurs concurrently with the accretion shock expansion, and may be associated with the draining of the inner disk into the black hole and with a simultaneous readjustiment of the pressure-supported atmosphere of the disk toward near-adiabatic stratification in the presence of convection or large scale circulation.  This decline in disk density can explain $d\ln \dot M/d\ln t\approx -2$ but not steeper.  

Second, there is a very gradual decrease, by a factor of $\lesssim 2$, of the vertical pressure scale height of the rotationally supported disk during the period of the steep decline.  The decrease can be seen in a comparison of the left panel of Figure \ref{fig:density}, showing the density distribution at $t=100~\textrm{s}$, with the right panel of the same figure, showing the density at $t=1,000~\textrm{ s}$.  Since in the rotationally supported flow the viscosity is proportional to the square of the scale height, the scale height decrease implies a factor of $\lesssim 4$ decrease of the viscosity $\nu$, and with it also of the disk accretion rate $\dot M_{\rm disk}$. Consistent with the disk scale height decrease, the midplane temperature of the disk decreases gradually and steadily.  E.g., in Run 2 at the innermost resolved radius of $10^8~\textrm{cm}$, the temperature drops from $5\times10^9~\textrm{K}$ at the onset of circularization to $2\times10^9~\textrm{K}$ at the end of Phase I.  

Third, there is a rapid decline of the rate at which the low angular momentum fluid accretes through the axial funnel.  Funnel accretion dominates the net accretion rate immediately following accretion shock formation but then drops to zero at the end of the steep decline at $t\sim 500~\textrm{s}$ when the funnel inflow reverses into an outflow.  Our simulations may overestimate the funnel accretion rate if the funnel material is additionally heated by a narrow relativistic axial jet, presumably launched from the black hole magnetosphere and responsible for the $\gamma$-ray and X-ray emission, that we do not simulate, but which must pierce the funnel region \citep[see, e.g.,][]{RamirezRuiz:02,Zhang:03,Zhang:04,ZhangW:06,Morsony:07,Wang:08}.  The effect of the heating of the funnel fluid by the relativistic jet might be to further steepen the decline of the accretion rate. Further magnetic outflow could develop from the corrona inner accretion disk, which could shut off funnel accretion more effectively than the outflow driven thermally by the resistive (or, in our approximation, viscous) dissipation in the disk \citep{Proga:03b,Proga:03c}.  These effects could clearly make the central accretion rate decline, which is already rapid in our simulations, become even more rapid.

One also expects that the accretion rate decline is accompanied by an inward recession of the boundary separating the radiatively-efficient, neutrino-dominated accretion flow (NDAF) and the radiatively-inefficent, advection or convection dominated accretion flow (RIAF) \citep[see, e.g.,][]{Chen:07}.  The evolution of an NDAF into an RIAF at radii $r\lesssim 10^8~\textrm{cm}$ over the course of a few hundred seconds is a process that may further accelerate the accretion rate decline in real GRB progenitors.  The physics of the transition from NDAF to RIAF and the onset of the outward propagation of the accretion shock are closely linked---both are controlled by neutrino cooling. We will argue in Section \ref{sec:triggering_decline} below that the two operating together, starting at about the same time, is the most likely reason for the rather steep decline of the accretion rate between several tens of seconds and several hundred seconds.

The rapid decline of the central accretion rate in our simulations is distinct in origin from the less rapid decline seen in the simulations of \citet{MacFadyen:01}.  MacFadyen et al.\ simulated the fallback of the stellar envelope following the failure of the shockwave resulting from the core bounce to unbind the star.  Placing their inner numerical boundary at $r_{\rm min,MHW}=10^9~\textrm{cm}$, they found that the radial fallback rate through the inner boundary declines at the rate $\dot M(r_{\rm min,MHW})\propto t^{-5/3}$.  Since the boundary was place outside the radii of the infalling envelope encounters the centrifugal barrier, MacFadyen et al.\ did not simulate the accretion disk and thus did not observe the formation and outward propagation of the accretion shock.   In our simulations, the accretion shock is aided by the viscous energy deposition in the rotationally-supported disk.  The post-circularization shock seems to be responsible for the much more rapid central accretion rate decline in our simulations than in those of MacFadyen et al.

The rapid temporal central accretion rate variability evident in Figure \ref{fig:mdot}\emph{c} is an outcome of hydrodynamical instabilities near the innermost simulated radius $R_{\rm min}$ (see, also, \citealt{MacFadyen:99}, who observed similar variability in Phase 0) and should not translate into any potential variability of the electromagnetic jet launched from radii $R\ll R_{\rm min}$.  The nature of the variability in the inward-directed mass flux $\dot M$ should also be affected by the fluctuations of the magnetic stresses \citep[e.g.,][]{Proga:03b,Proga:03c} and by the complex interplay of the processes associated with nuclear reactions and neutrino transport in the accretion flow \citep[e.g.,][]{MacFadyen:99,Proga:03c,Nagataki:07,LopezCamara:09}. Therefore, we caution against ascribing phenomenological significance to the accretion rate variability in Figure \ref{fig:mdot}\emph{c}.

\begin{figure}[t]
\begin{center}
\includegraphics[width=0.45\textwidth,clip]{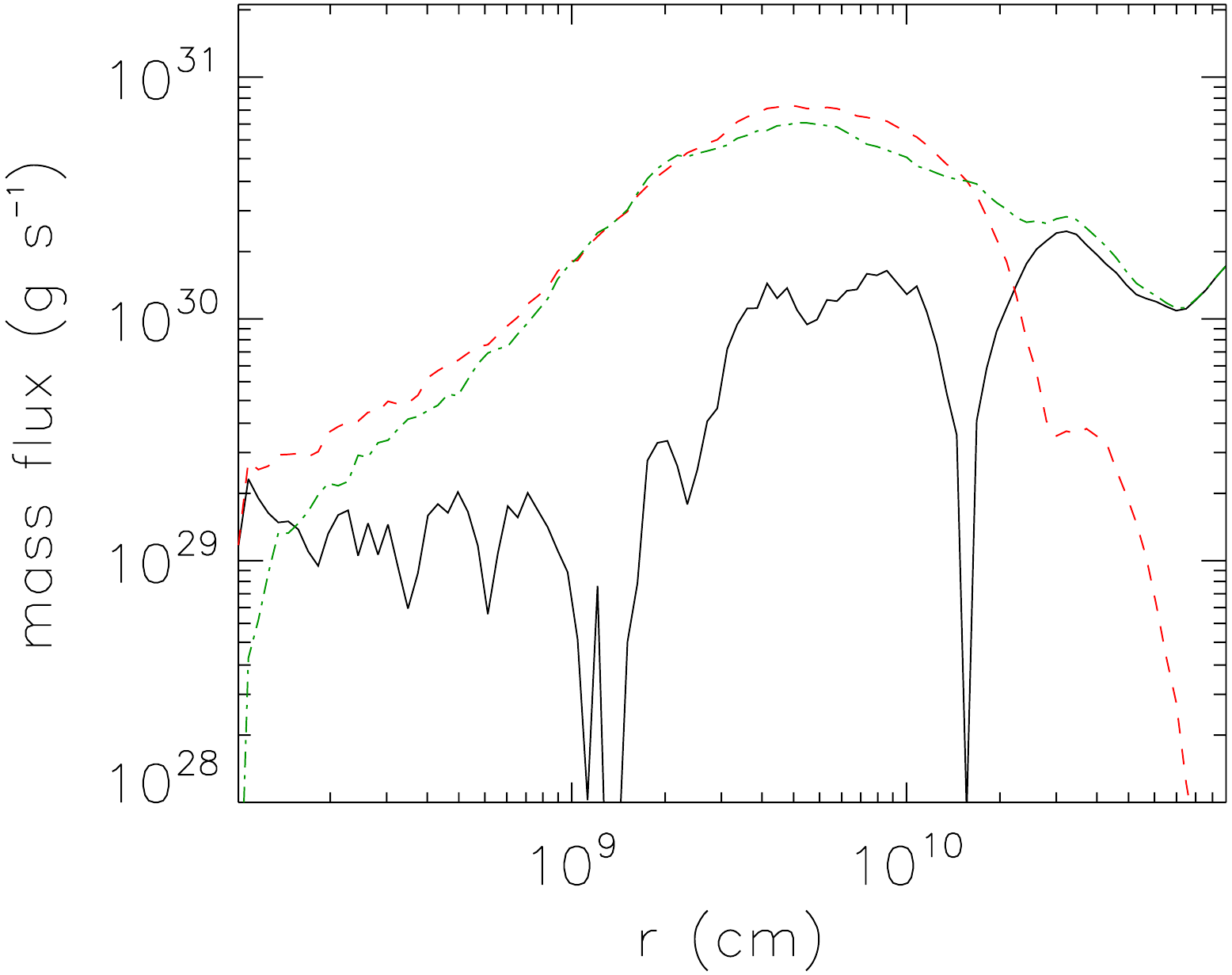}
\end{center}
\caption{The rate of mass inflow (\emph{red dashed line}), mass outflow (\emph{green dot-dashed line}), and absolute net mass flow (\emph{black solid line}) crossing a sphere of radius $r$ centered on the black hole, in Run 2.  The rates were averaged over the time interval $t=600-1,000~\textrm{s}$.  The outflow and the inflow nearly cancel over a range of radii.  At the radii of the rotationally-supported disk $r\lesssim 10^9~\textrm{cm}$, there is net inflow at the rate $\dot M\sim 5\times10^{-5}~M_\odot~\textrm{s}^{-1}$.}
\label{fig:mass_flux}
\end{figure}

\begin{figure*}[t]
\begin{center}
\includegraphics[width=0.45\textwidth,clip]{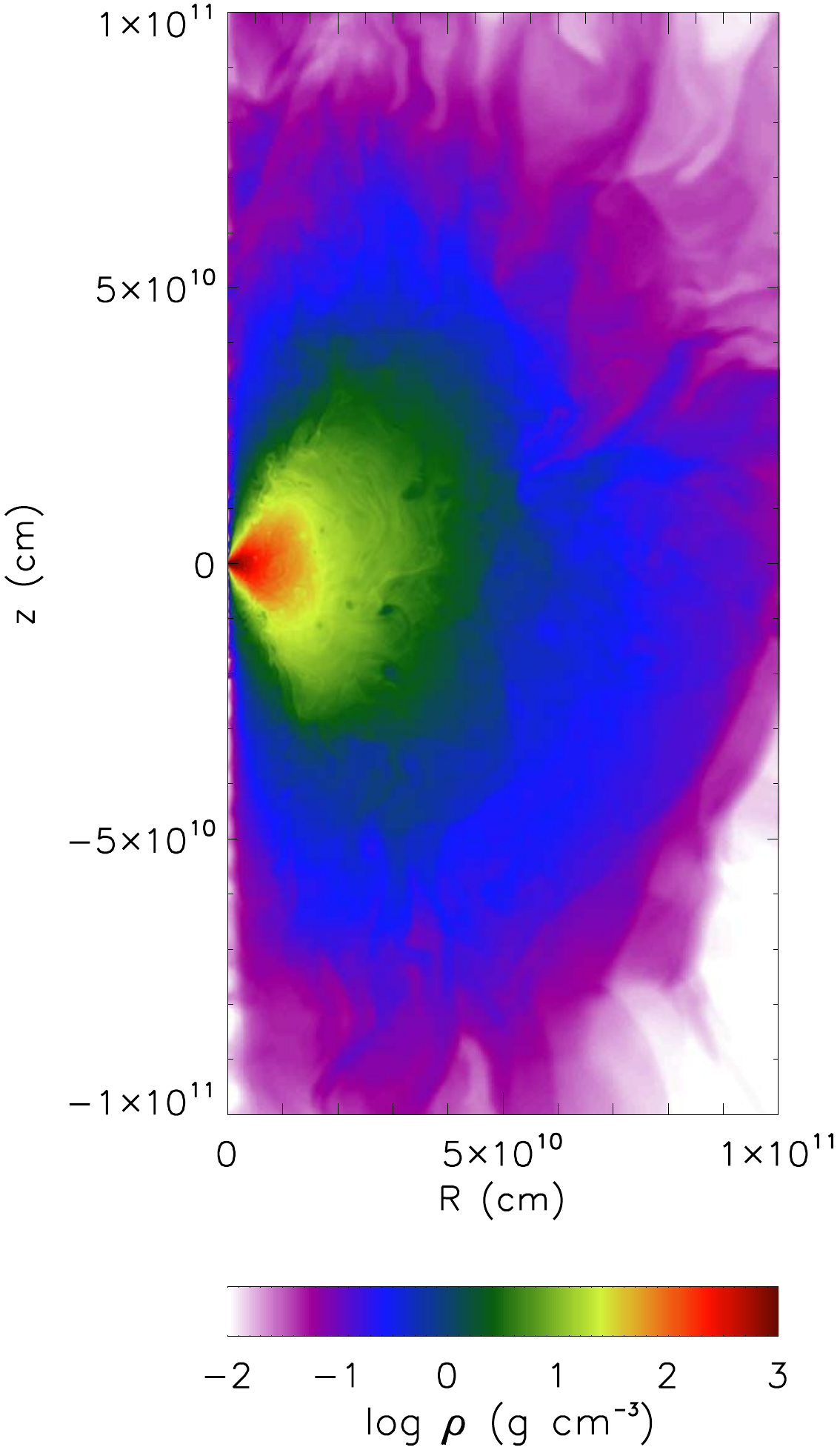}
\includegraphics[width=0.45\textwidth,clip]{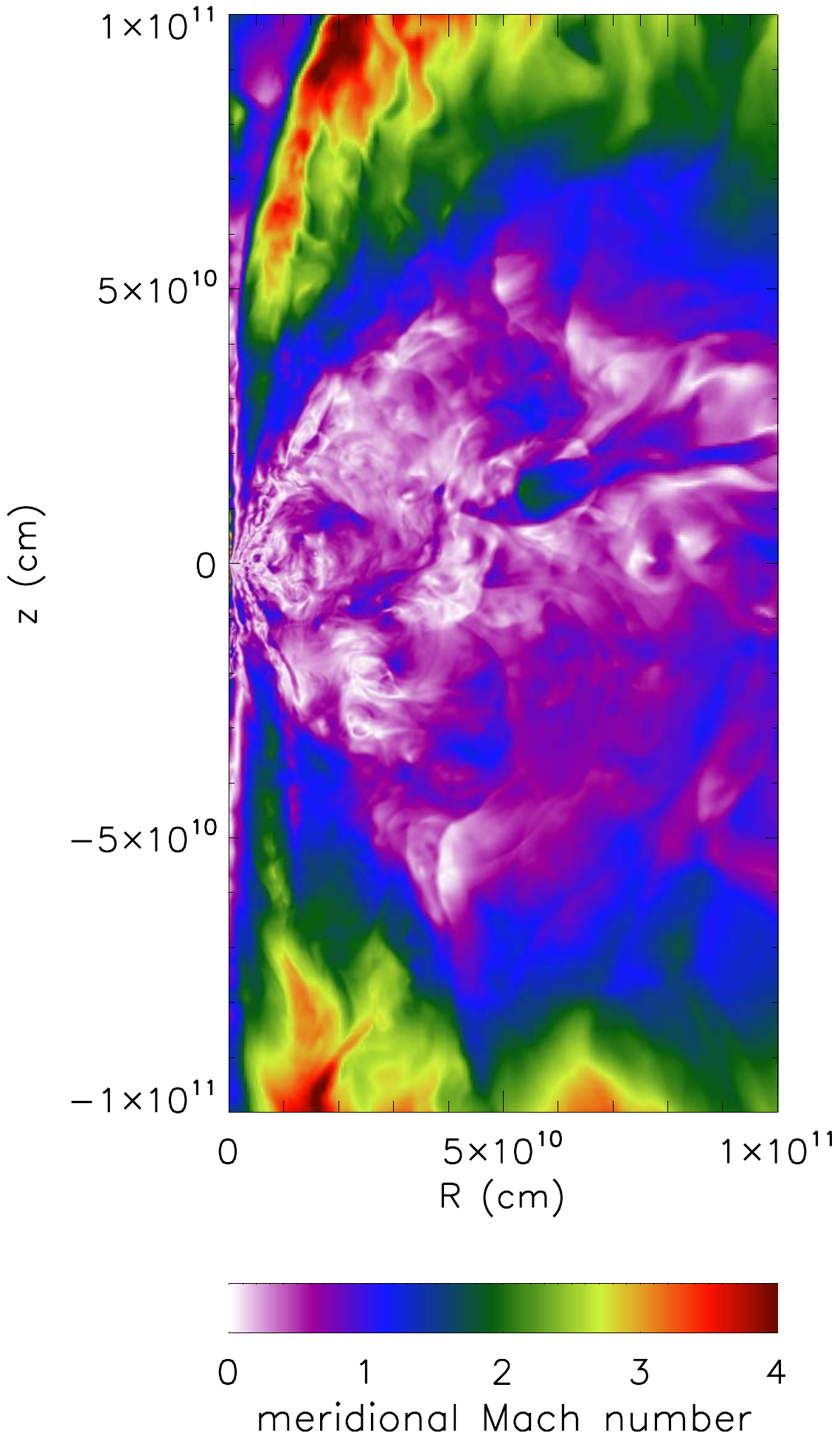}
\end{center}
\caption{The large-scale density distribution (\emph{left panel}) and meridional Mach number $M_{R-z}\equiv (v_R^2+v_z^2)^{1/2}/c_{\rm s}$, where $c_{\rm s}$ is the adiabatic sound speed (\emph{right panel}) at $t=2,000~\textrm{s}$ in the lowest-resolution simulation, Run 3. Meridional motions in the pressure-supported atmosphere that contains most of the unaccreted mass are subsonic, indicating that large-scale infall has ceased.  The supersonic fluid has positive Bernoulli constant and is unbound (see Figure \ref{fig:bernoulli}).}
\label{fig:whole_star}
\end{figure*}

\subsection{Phase II: Funnel Outflow, Thick Disk Accretion}
\label{sec:phase_2}

The steep decline of the accretion rate seems to diminish, or even cease at $\sim 200-500~\textrm{s}$, and the accretion rate seems to transition to a quasi-steady regime.  This behavior resembles the transition into the ``plateau'' phase, or Phase II, of the GRB X-ray light curve.  The simulated accretion flow appears to settle in a quasi-steady state, characterized by an axial outflow and thick equatorial disk accretion.  We proceed to characterize the quasi-steady accretion flow.  Figure \ref{fig:acceleration}, which shows the magnitudes of the various terms in the spherically-averaged Euler equation and the net residual acceleration implied by the radial Euler equation, indicates that bulk of the fluid mass is rotationally supported at $r\lesssim 3\times10^9~\textrm{cm}$ and is pressure supported at $r\gtrsim 4\times10^9~\textrm{cm}$.  The relative contribution of pressure support at radii where rotational support dominates is still substantial, $\sim 50-75\%$, consistent with the thick disk morphology with vertical scale height $h/R\sim \tan 30^\circ$ (Figure \ref{fig:density}, right panel).  Thus, our post-core-collapse accretion flow never resembles a thin disk. The pressure-supported atmosphere is nearly isentropic, $p\propto \rho^\gamma$.

In Figure \ref{fig:mass_flux}, we show the inward-directed, the outward directed, and the net mass flux flowing through spherical shells with radii $r$.  After $\sim 500\textrm{ s}$, the net mass flux in the central $\sim 10^9~\textrm{cm}$ is approximately independent of radius, which reflects a quasi-steady accretion in the inner part of the rotationally-supported disk at the rate $\dot M_{\rm disk}\approx 5\times10^{-5} ~M_\odot~\textrm{s}^{-1}$.  This disk accreting in a quasi-steady state contains only $M_{\rm disk}\sim 0.01~M_\odot$, which is less than $1\%$ of the mass that remains bound to the black hole in the shock-heated, pressure-supported atmosphere atop the rotationally-supported disk.  The outflow and the inflow nearly cancel over the range of radii belonging to the atmosphere, just as was found in the simulated radiatively inefficient accretion flow with convection of \citet{Abramowicz:02}.  The structure described by a massive convective atmosphere surrounding a thick, nonradiative disk resembles the ``quasistar'' of \citet{Begelman:08}, who envisioned the limit in which the mass of the pressure supported envelope exceeds the mass of the black hole by a large factor.

The accretion time of the inner disk during Phase II, $t_{\rm acc}\sim M_{\rm disk}/\dot M_{\rm disk}\sim  200~\textrm{s}$, is shorter than duration of this phase (we end our simulation prior to the end of Phase II), hence a continuous replenishment of the inner disk must operate.   The time scale on which the entire fluid mass bound to the black hole ($\sim 2M_\odot$) would accrete through the thick disk is $\sim 4\times10^4~\textrm{s}$, though of course, not all of the mass bound at the beginning of Phase II must ultimately accrete; a large fraction could become unbound and leave in an outflow.  Because of computational limitations we do not extend the simulations long enough to observe the inevitable depletion of the massive atmosphere through inner disk accretion, but in Section \ref{sec:long_term}, we speculatively extrapolate our results into that regime.

In Figure \ref{fig:whole_star}, we show a large-scale ($\sim 10^{11}~\textrm{cm}$) view of the density and the meridional Mach number $M_{R-z}\equiv (v_R^2+v_z^2)^{1/2}/c_{\rm s}$, where $c_{\rm s}$ is the adiabatic sound speed at $t=2,000~\textrm{s}$ in the lowest-resolution simulation, Run 3. Meridional motions in the pressure-supported atmosphere are subsonic, confirming that large-scale infall has ceased.
Because a vast fraction of the unaccreted mass is in this atmosphere, we neither observe nor anticipate the tendency of the inner disk to spread outward in the way in which an isolated thin disk would spread and how \citet{Cannizzo:09} envision.  At the quasi-steady disk radii, the inner and outward-directed mass fluxes increase outward according to $\dot M_{\rm in}(r), ~ \dot M_{\rm out}(r) \propto r^{1.0} - r^{1.2}$, which reflects the convective or circulatory nature of the flow.  The pressure-supported atmosphere at radii $5\times10^9~\textrm{cm} \lesssim r \lesssim 10^{10} ~\textrm{cm}$ contains about $0.5~M_\odot$ and exhibits a net inflow at the rate $\approx 5\times10^{-4}~M_\odot~\textrm{s}^{-1}$,  larger than in the inner disk; the lack of a true steady state opens the prospect for a late-time, high-amplitude central accretion rate variability.  On the other hand, the outer atmosphere $r\gtrsim 2\times10^{10}~\textrm{cm}$ containing $\sim 2~M_\odot$ has a net outward-directed mass flux at the rate $\dot M_{\rm out}\sim (0.5-1.5)\times10^{-3}M_\odot~\textrm{s}^{-1}$, though most of the outflowing mass remains gravitationally bound to the black hole.

\begin{figure}[t]
\begin{center}
\includegraphics[width=0.45\textwidth,clip]{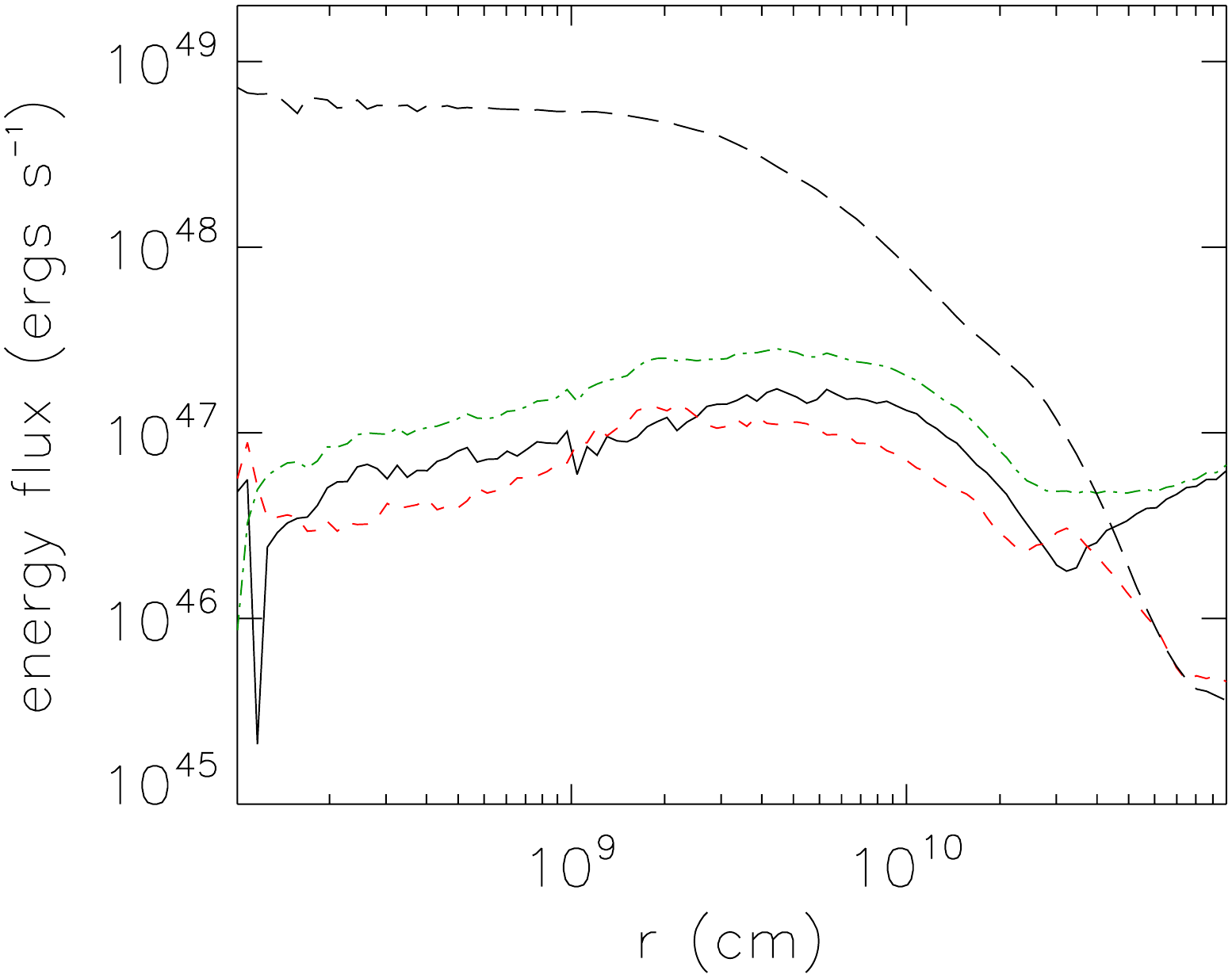}
\end{center}
\caption{The rate of energy inflow (\emph{red short dashed line}), energy outflow (\emph{green dot-dashed line}), and absolute net energy flow (\emph{black solid line}) crossing a sphere of radius $r$ centered on the black hole in Run 2.  The rates were averaged over the time interval $t=600-1,000~\textrm{s}$.  Energy outflow dominates inflow at all radii. The black long dashed line shows the product of pressure and the sound speed to which the maximum energy that can be transported by convection is proportional.}
\label{fig:energy_flux}
\end{figure}

Figure \ref{fig:energy_flux} shows that the entire rotationally supported region $2\times10^8 ~\textrm{cm}\lesssim r\lesssim 5\times10^9~\textrm{cm}$ exhibits a net outward-directed energy flux 
\beq
\dot E (r) \sim 10^{47} ~\textrm{erg s}^{-1} \left(\frac{r}{2\times10^9~\textrm{cm}}\right)^{0.4} ,
\eeq
which in the steady state disk, $r\lesssim 10^9~\textrm{cm}$, implies a mass conversion efficiency of $\dot E/(\dot M c^2)\sim 7\times10^{-4}$.  In the innermost cells $r\sim 10^8~\textrm{cm}$, however, there is a hint of an energy inflow.

\begin{figure*}
\begin{center}
\includegraphics[width=0.45\textwidth,clip]{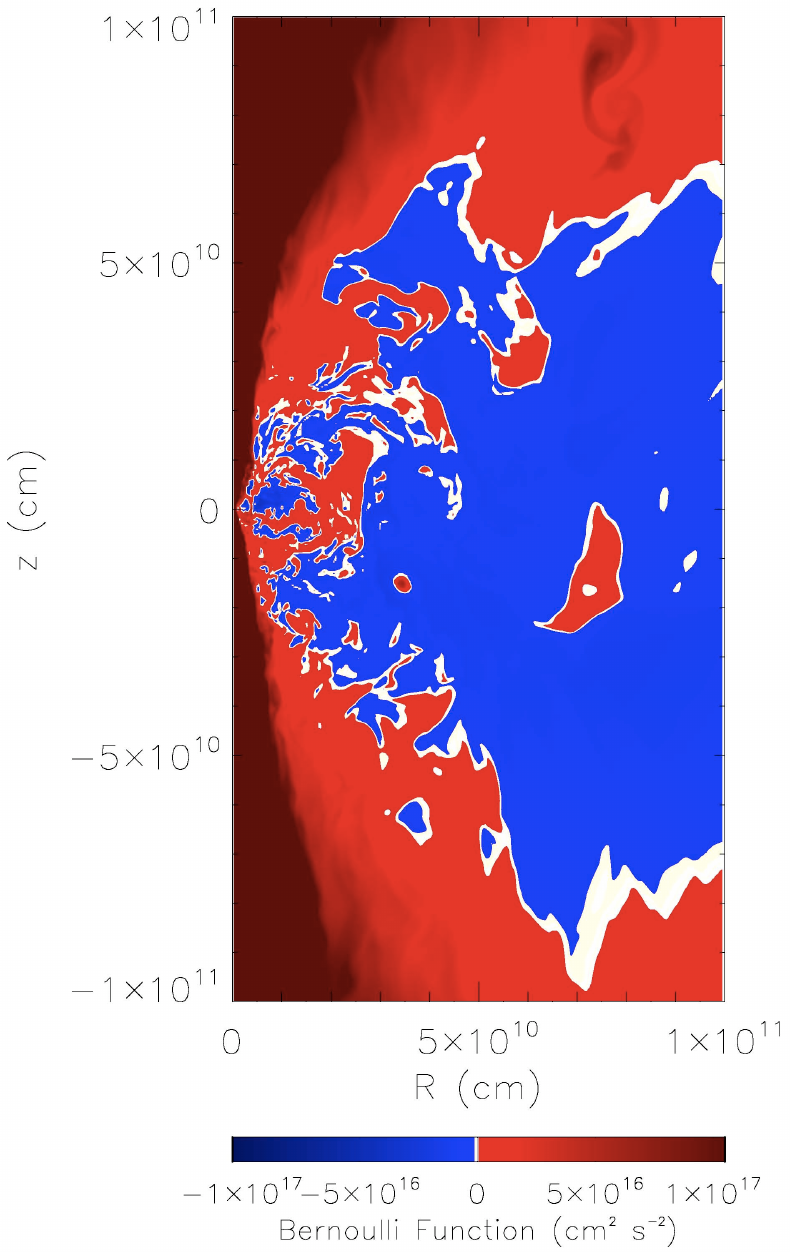}
\includegraphics[width=0.45\textwidth,clip]{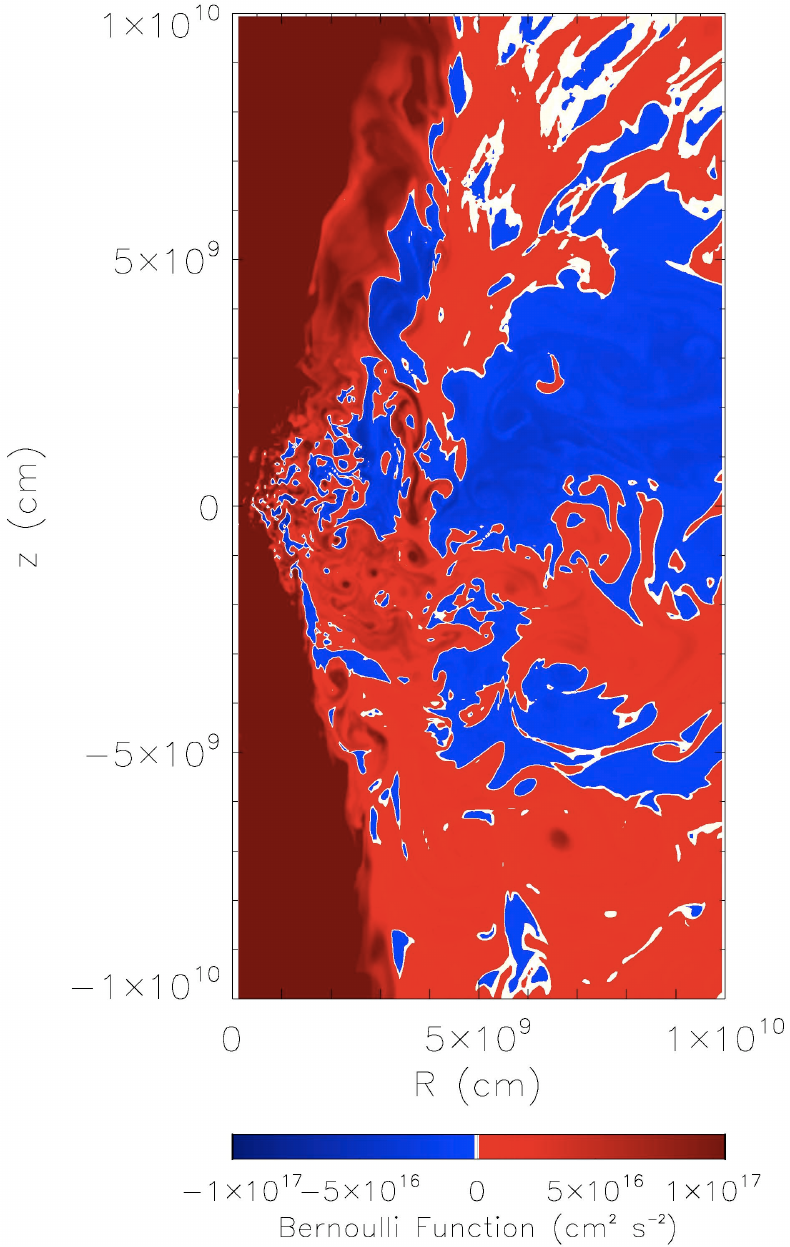}
\end{center}
\caption{Value of the Bernoulli function $Be\equiv E_{\rm k}+E_{\rm p}+[\gamma/(\gamma-1)]~p/\rho$, where $E_{\rm k}$ and $E_{\rm P}$ denote, respectively, the specific kinetic and the potential energy, in the simulation (\emph{left panel}) and in the central $10^{10}~\textrm{cm}$ (\emph{right panel}) at $t=1,000~\textrm{s}$ in Run 2. The material with positive values of the Bernoulli function (\emph{red color}) has enough energy to escape from the system.}
\label{fig:bernoulli}
\end{figure*}

To search for the presence of unbound flows, in Figure \ref{fig:bernoulli}, we plot the Bernoulli function, defined as the sum of the specific kinetic energy, enthalpy, and potential energy of the accretion flow \citep[e.g.,][]{Narayan:94,Narayan:95,Stone:99,Igumenshchev:99,Igumenshchev:00a,Blandford:99,Blandford:04}\footnote{Some authors also define the Bernoulli constant $b$ to equal the Bernoulli function divided by the Keplerian velocity, $b\equiv Be/v_{\rm K}^2$ \citep{Narayan:94,Narayan:95}.} 
\beq
Be \equiv \frac{1}{2} (v_R^2 + v_z^2 + v_\phi^2) + \frac{\gamma}{(\gamma-1)}~\frac{p}{\rho} + \Phi ,
\eeq
where $\Phi$ is the total, negative gravitational potential.  The structure at small radii bears only a coarse-grained resemblance to the Bernoulli function profile in the simulations of \citet{Stone:99}.  The entire axial funnel region, plus scattered domains within the equatorial rotationally supported and pressure supported region, as well as the high-latitude fluid, $|z|\gtrsim 7\times10^{10}~\textrm{cm}$, exhibit a positive Bernoulli function, $Be>0$, which indicates the potential for an unbound flow and an escape to infinity. The positivity of the Bernoulli function does not always necessitate escape, as such fluid elements can be buried within the massive quasi-hydrostatic envelope where they can interact and mix with the $Be<0$ fluid.  The vertically outflowing fluid above and below the rotationally supported disk, within an angle of $\sim 20^\circ-25^\circ$ from the vertical axis, however, is entirely unbound, in clear indication of the presence of mass loss carried by a thermally-driven disk wind.  Indeed, the axial outflow becomes supersonic as it propagates upward through the envelope (see Fig.\ \ref{fig:whole_star}).  Therefore, the system spontaneously develops an advection-dominated inflow-outflow solution (ADIOS; \citealt{Blandford:99,Blandford:04}).

\section{Discussion}
\label{sec:discussion}

The primary limitations of the numerical model presented here include: (i.) the lack of simulation coverage of the hot inner disk, $r\lesssim 5\times10^7~\textrm{cm}$, where neutrino and nuclear physics influence the thermodynamics of the flow, (ii.) the limited adequacy of the Navier-Stokes viscous fluid dynamics to approximate the dynamics of a realistic magnetized, radiation dominated fluid, (iii.) the lack of modeling of the axial relativistic jet and its enveloping cocoon, and (iv.) the lack of coverage of the very late evolution, $t\gtrsim 10^4~\textrm{s}$, when the GRB X-ray light curve exhibits single or multiple breaks with the tendency toward a steepening of the luminosity decline.  We defer an exploration of the limitations (ii.) and (iii.) for future work, and here briefly and speculatively address limitations (i.) and (iv.).  In Section \ref{sec:triggering_decline}, we crudely take into account the energy loss to neutrino emission in the accretion-shock-heated flow and estimate the time at which the central accretion rate commences to decline steeply (the transition from Phase 0 to Phase I).  In Section \ref{sec:long_term}, we discuss the implications of mass loss for the long-term evolution of accretion rate.

\subsection{Triggering of the Steep Decline in GRBs}
\label{sec:triggering_decline}

The composite $\gamma$-ray and X-ray GRB light curve starts declining abruptly and steeply after an initial period of steady luminosity that lasts for tens of seconds.  Here, we attempt to shed light on the transition from the quasi-steady activity of the central engine to the steeply declining regime.  Continuing to work within the paradigm in which the luminosity is proportional to the central accretion rate, we suggest that the steep decline is triggered by a rapid outward expansion of an accretion shock through the infalling material that feeds a convective, rotationally-supported thick accretion disk.  This is consistent with the conclusion of \citet{BarniolDuran:08}, who ruled out mechanisms for powering the rapid decline that are intrinsic to a single emitting and cooling element, and by elimination inferred that the central engine (e.g., an accreting black hole) remains active during the steeply declining phase (but see \citealt{Genet:08}, who showed that the high latitude emission from a sequence of such elements, or ``pulses,'' could fit the decline). Because in our adiabatic simulations we do not resolve the innermost radii $r\lesssim 5\times10^7~\textrm{cm}$ and do not simulate the nuclear photodisintegration, neutrino emission, neutrino capture on nucleons, and neutrino annihilation, that occur in the hot ($T\gtrsim10^{10}~\textrm{K}$) plasma at these radii, some of the forthcoming conclusions will be obtained with the aid of a simple analytical model for the thermodynamic evolution of the inner neutrino-cooling region.

We have seen that in our adiabatic simulations, the accretion shock appears when the infalling material has enough angular momentum to be rotationally supported at the innermost resolved radius.  From there on, the accretion shock expands outward rapidly and sweeps through the star.  The shocked fluid is additionally heated as a result of the viscous dissipation in the thick disk.  The energy produced during the accretion of the thick disk is advected or convected radially outward by the disk ``wind'' (Figure \ref{fig:entropy}) which distributes it throughout the mostly pressure-supported torus of shocked fluid.  In the adiabatic regime, after the gravitational potential becomes dominated by the black hole while the mass infall rate declines rapidly, it seems that an outward expansion of the hot torus bounded by the accretion shock is inevitable immediately following shock formation.  In the nonadiabatic regime, the expansion of the shock may be delayed by losses associated with nuclear photodisintegration and neutrino emission. If indeed, in general, a sudden and rapid drop in the central accretion rate accompanies the shock expansion, then to estimate the onset of the steep decline, one must identify the instance at which the losses become inefficient and the accretion shock, aided by the viscous energy injection, can start traveling outward.

Consider a fluid element with specific angular momentum $\ell=10^{17}~\ell_{17}~\textrm{cm}^2~\textrm{s}^{-1}$; its circularization radius around the black hole of mass $M_{\rm BH}=10~M_1~M_\odot$ is 
\beq
r_{\rm circ}\sim \frac{\ell^2}{G M_{\rm BH}} . 
\eeq 
If the material with density $\rho =10^8~\rho_8~\textrm{g cm}^{-3}$  arrives at the shock from a free fall from infinity and the gravitational energy density $GM_{\rm BH} \rho /r_{\rm circ}$ is converted into the energy density in radiation $a T^4$, where $a$ is the radiation constant, the post-shock temperature is given by 
\bea
\label{eq:T_shock}
T_{\rm shock} &\sim& \left(\frac{GM_{\rm BH}}{\ell}\right)^{1/2} \left(\frac{7~\rho}{a}\right)^{1/4}\nonumber\\
&\sim& 6.4\times10^{10} ~\textrm{K}~\frac{M_{1}^{1/2} ~ \rho_{8}^{1/4}}{\ell_{17}^{1/2}} ,
\eea
where we have taken the density jump across the shock to be $\sim 7$, appropriate if the fluid velocity is Newtonian and the post-shock fluid is radiation-dominated.  The latter in particular is a crude approximation; \citet{Chen:07} show that the pressure due to baryons and electrons and positrons can be comparable to and larger than the radiation pressure in the disk midplane.

\citet{Nagataki:07} find that almost all of the energy emitted by neutrinos in the hot, rotationally-supported torus, comes from pair capture on free nucleons (the Urca process),\footnote{In the calculation of \citet{LopezCamara:09}, the cooling due to neutrino emission from pair annihilation dominates at early times $t=0.2\textrm{ s}$ and radii $R\lesssim 2\times10^7~\textrm{ cm}$ on the equatorial plane; at later times, the cooling from pair annihilation is comparable to the cooling from pair capture.} for which the approximate cooling rate is \citep[and references therein]{Qian:96,Popham:99}
\beq
\label{eq:q_eN}
q_{eN} = 9\times10^{25} ~\rho_8 ~T_{10}^6~ X_{\rm nuc} ~\textrm{erg cm}^{-3} ~\textrm{s}^{-1} ,
\eeq
where $T=10^{10} ~T_{10} ~\textrm{K}$ is the plasma temperature, and $X_{\rm nuc}$ is the free nucleon fraction in statistical equilibrium, which is approximated via
\beq
\label{eq:X_nuc}
X_{\rm nuc} \approx {\rm min} [1,~8.7~ \rho_8^{-3/4} ~T_{\rm 10}^{9/8} ~\exp(-6.1/T_{\rm 10})] .
\eeq
While the cooling time 
\beq
\label{eq:t_cool}
t_{\rm cool} = \frac{a T^4}{q} ,
\eeq
where $q$ is the net energy loss rate, which includes the pair capture term $q_{eN}$ and other contributions, is shorter than the age of the collapse $t$, the accretion shock is confined near the black hole, the flow crossing the shock is highly supersonic, and a high accretion rate onto the central object is possible.  

When the cooling time exceeds the age of the collapse, $t_{\rm cool}>t$, the accretion shock expands outward. It seems that in the specific case of the collapse of a 16TI-model star, the pair capture neutrino emission indeed dominates losses until the cooling is no longer able to prevent outward expansion of the shock.  When the post-shock temperature drops below $\sim 9
\times 10^{9}~\textrm{K}$, which happens due to the decrease in the free nucleon abundance under the conditions of nuclear statistical equilibrium, the losses from nuclear photodisintegration and from neutrino emission from pair annihilation exceed those from pair capture, but they are not able to prevent shock expansion. 

Consider an initial stellar density profile of the form $\rho \propto r^{-\tau}$ in which the gravity is dominated by the mass closer to the center, and let 
\beq
M(r)=M_0\left(\frac{r}{r_0}\right)^{3-\tau}
\eeq
denote the pre-collapse mass profile, where $M_0$ and $r_0$ denote the stellar mass and radius, respectively (we ignore the departure of the density profile from a single power law near the stellar surface).  The free fall time from radius $r$ is given by $t_{\rm ff}(r) \sim [r^3/GM(r)]^{1/2}$.   This relation can be inverted to obtain the radius, defined via $t_{\rm ff}(r_{\rm ff})=t$, from which the freely falling material is reaching the center at time $t$, 
\beq
r_{\rm ff} (t) = (G M_0 t^2 r_0^{\tau-3})^{1/\tau} .
\eeq
The mass of the black hole grows in time and approximately equals
\bea
M_{\rm BH} (t) &\sim& M[r_{\rm ff}(t)]\nonumber\\ 
&\sim& G^{3/\tau-1} ~M_0^{3/\tau} ~r_0^{3(1-3/\tau)} ~t^{2(3/\tau-1)} 
\eea
Assuming that $r_{\rm circ}\ll r_{\rm ff}$, the pre-shock density of infalling fluid at $r_{\rm circ}$ at time $t$ since the beginning of the explosion approximately equals the mass infall rate $\sim M[r_{\rm ff}(t)]/t$ divided by the shock area $\sim 4\pi r_{\rm circ}^2$ multiplied by the infall velocity $\sim (GM_{\rm BH}/r_{\rm circ})^{1/2}$,
\bea
\label{eq:rho_pre_shock}
\rho(t) &\sim& \frac{M[r_{\rm ff}(t)]}{4\pi [G M_{\rm BH}(t)~r_{\rm circ}^3]^{1/2} t}\nonumber\\
&\sim& \frac{G^{1-6/\tau} ~M_0^{6/\tau}~ r_0^{6(1-3/\tau)} ~t^{12/\tau-5}}{4\pi ~\ell(t)^3} .
\eea
We further assume that 
\beq
\label{eq:ell_profile}
\ell(r) \sim \ell_0 ~ \left[\frac{M(r)}{M_0}\right]^{\eta} ~\textrm{cm}^2~\textrm{s}^{-1} ,
\eeq  
implying that 
\beq
\ell(t) \sim  \ell_0 ~(GM_0)^{(3/\tau-1)\eta} ~r_0^{3(1-3/\tau)\eta} ~t^{2(3/\tau-1)\eta} .
\eeq

\citet{Kumar:08b} find that $\tau \approx 2.5$ throughout the bulk of the star for the model 16TI of \citet{Woosley:06a} that we utilize and we adopt this value. We further find that $\eta\approx 2.5$ is consistent with the rotational profile of the model 16TI in the range $4~M_\odot\lesssim M(r) \lesssim 10~M_\odot$.   We set $r_0=10^{10}~\textrm{cm}$, and $M_0=10~M_\odot$ and $\ell_0=10^{17.8}~\textrm{cm}^2\textrm{ s}^{-1}$, which approximate the mass and angular profile of the model 16TI, and substitute $r=r(t)$ in equation (\ref{eq:ell_profile}) and substitute $\ell[r(t)]$ in equation (\ref{eq:rho_pre_shock}) to obtain
\bea
\label{eq:rho_pre_shock_16TI}
\rho &\sim& \frac{M_0^{9/10}~r_0^{33/10}}{4\pi~G^{1/10} \ell_0^3 ~t^{16/5}} \nonumber\\
&\sim& 10^{10}~ t^{-16/5} ~\textrm{g cm}^{-3}  ,
\eea
where in the last expression, $t$ is given in seconds.

Combining equations (\ref{eq:T_shock}), (\ref{eq:q_eN}), (\ref{eq:X_nuc}), (\ref{eq:t_cool}), and (\ref{eq:rho_pre_shock_16TI}), the ratio of the cooling time for pair capture only, $q=q_{eN}$, to the age of the collapse reads
\beq
\label{eq:cooling_time_ratio}
\frac{t_{\rm cool}}{t} \sim \frac{2\times10^{-6} ~t^{22/5}}{{\rm min}[1,17.5~t^{93/80}~\exp(-0.27~t^{11/10})]} .
\eeq
The ratio rises rapidly in time and becomes unity $t_{\rm cool}/t\sim 1$ at 
\beq
t_{\rm decl}\approx 20~\textrm{s} .
\eeq
At this point, the material circularizing at $r_{\rm circ}$ is no longer able to cool by neutrino emission.  Therefore, we expect that in a realistic star corresponding to the model 16TI, the accretion rate starts to decline steeply at $t_{\rm decl} \sim 20~\textrm{s}$ after the explosion, when the mass of the black hole is $\sim 9M_\odot$ and post-shock temperature is $\lesssim 10^{10}~\textrm{K}$.  Since the initial mass of the black hole, if taken to equal the mass of the iron core, is $M_{\rm BH,init}\sim 1.5~M_\odot$, the implied average accretion rate preceding the decline, $\langle \dot M \rangle = [M_{\rm BH}(t_{\rm decl}) - M_{\rm bh,init}]/t_{\rm decl} \sim 0.4 ~ M_\odot ~\textrm{s}^{-1}$, is larger than the accretion rate $\sim (0.1-0.2) ~M_\odot~\textrm{s}^{-1}$ observed in our simulations and those of \citet{MacFadyen:99} and \citet{Nagataki:07}, though it is consistent with the accretion rate in the first $0.3~\textrm{s}$ in the simulation of \citet{Proga:03c} and in the first $0.4~\textrm{s}$ in the simulation of \citet{LopezCamara:09}.  Our model possibly overestimates the infall rate as it does not take into account the initial hydrostatic pressure gradients that delay the collapse in the simulations and in real GRB progenitors.  

If this model for the triggering of the steep decline of the central accretion rate is correct, and if the onset of the steep decline of the accretion rate implies a termination of the observable prompt $\gamma$-gray emission, then the duration of the prompt emission in long GRBs produced by black hole-forming core collapse events should be anti-correlated with the angular momentum of the progenitor envelope.  High angular momentum envelopes circularize at large radii where low virial temperatures imply post-shock adiabaticity and an earlier heating of the infalling envelope by the outward expanding accretion shock and disk outflows.  Low angular momentum envelopes may circularize at radii where the high virial temperatures imply rapid cooling over several tens of seconds after the initial collapse.  The cooling allows the accretion shock to remain confined longer at radii where the free-fall velocity of the infalling envelope is highly supersonic and a high central accretion rate is possible, barring, of course, another process, such the electromangetically-driven outflow observed in, e.g., \citet{Proga:03c}, \citet{Nagataki:07}, and \citet{Komissarov:09}, that could suppress central accretion.

\subsection{The Long-Term Evolution}
\label{sec:long_term}

The period of quasi-steady or gradually declining luminosity in GRB X-ray light curves lasts for $\sim10^3-10^4~\textrm{s}$ (Phase II).  At the end of this period, a steeper decline resumes (Phase III), but with the shallower slope  $L_{\rm X}\propto t^{-1.2}$ than in the steeply declining regime of Phase I.  Occasionally, at $t\sim 10^4-10^5~\textrm{s}$, an even steeper decline $L_{\rm X}\propto t^{-2}$ takes over (Phase IV in the nomenclature of \citealt{ZhangB:06}).  If the steepening of the light curve reflects an underlying decline of the central accretion rate, what process is responsible for this decline?  Possibilities include a transformation of the character of the accretion flow due to an internal redistribution of material inside the accreting envelope and a depletion of the mass reservoir that feeds the central accretion. 

At the densities of the accreting, pressure-supported envelope, which are $>10^{-3}~\textrm{g}~\textrm{cm}^{-3}$ at the outer boundary of the simulation box at the end of each run, the radiation is effectively trapped and internal radiation transfer in the disk and the envelope is not important on the time scales on which X-ray light curve data are available.  It also seems that given the near-hydrodynamic equilibrium state at $t\sim10^3~\textrm{s}$, any longer-term internal hydrodynamic redistribution of material between the disk and the envelope should be very gradual, so such a hydrodynamic redistribution is probably not a candidate for the steepening that marks the transition from Phase II to Phase III or that which marks the transition from Phase III to Phase IV.  Depletion of the reservoir consisting of the rotationally supported disk and the pressure supported atmosphere could occur due to the accretion of the fluid into the black hole (``draining''), due to a hydrodynamic outflow launched from the surface of the thick disk and escaping through the axial funnel region (``venting''), and due to a radiatively-driven mass loss in the photosphere of the envelope (``blowoff'').

In the absence of mass loss to unbound flows, the time scale on which the gravitationally bound envelope drains into the black hole, estimated from the bound envelope mass ($M_{\rm env}\sim 2~M_\odot$) and the accretion rate ($\dot M\sim 5\times10^{-5}~M_\odot~\textrm{s}^{-1}$, see Section \ref{sec:phase_2}) at $t=600-1,000~\textrm{s}$ in Run 2 is 
\beq
t_{\rm acc} \sim \frac{M_{\rm env}}{\dot M} \sim 4\times10^4~\textrm{s} .
\eeq
This time scale is somewhat longer but within uncertainties consistent with the time scale of the initial steepening of the light curve at the transition from Phase II to Phase III.  If the evolution of the envelope under draining is self-similar $\dot M\propto M_{\rm env}$, one might expect an exponential decline of the accretion rate; such a self-similarity, however, is not necessarily expected.

The positive Bernoulli function of the fluid in the region of the axial funnel in Figure \ref{fig:bernoulli} and the  mass influx and outflux that increases with radius in Figure \ref{fig:mass_flux} suggest the possibility that the dominant depletion may not be to accretion into the black hole, but instead 
to the loss exacted by the wind launched thermally from the surface of the thick, convective, rotationally supported disk.  The peak net outflow rate at $r\sim 3\times10^{10}~\textrm{cm}$ in Run 2 is $\dot M_{\rm out}\sim 10^{-3} ~M_\odot~\textrm{s}^{-1}$, which implies a short depletion time scale of $t_{\rm loss}\sim500~\textrm{s}$.  This time scale, however, is almost certainly an overestimate given that the high outflow rate may be a transient associated with the incomplete readjustment to the passage and breakout of the primary accretion shock.  It seems evident, however, that the draining into the black hole and the mass loss to hydrodynamically and thermally driven outflows from the surface of the thick disk and the massive envelope can provide explanations of the termination of quasi-steady accretion marking the Phase II to Phase III transition.

Convective energy transport in the massive, pressure-supported envelope can continue out to some critical radius where the convective motions become supersonic, resulting in shocks, or where radiation diffusion across the convective cells thwarts the convective instability. Outside of this radius, energy is transported either radiatively or by non-convective bulk motions.  Since the energy flux is a factor of $10^7-10^8$ above the Eddington limit, radiation pressure accelerates the fluid outward, resulting in a supersonic wind \citep[see, e.g.,][]{Shaviv:01,Owocki:04,vanMerle:08}.  The wind mass loss rate is limited by energy conservation $\onehalf \dot M_{\rm wind} v_{\rm esc}^2 \leq L$, and if the wind driving is radiative,  momentum conservation, $\dot M_{\rm wind} v_\infty \leq L/c$, where $v_{\rm esc}$ is the escape velocity from the critical radius, and  $v_\infty$ is the velocity of the wind at infinity.  To our best knowledge, the mechanics of mass loss in this extremely super-Eddington regime have not been explored. There is the possibility that the atmospheric mass loss occurs on a time scale compatible with the final steepening of the GRB X-ray light curve, at the Phase III to Phase IV transition.  Alternatively, as we have argued above, the Phase II to Phase III transition, and the Phase III to Phase IV transition, could both be caused by non-radiative losses (the draining into the black hole and venting in the axial funnel), but longer-term simulations are required to check this possibility.

\section{Conclusions}
\label{sec:conclusions}

We have conducted hydrodynamic simulations of the viscous post-core-collapse accretion of a rapidly rotating $\sim 14 ~M_\odot$ Wolf-Rayet star of \citet{Woosley:06a} onto the central black hole.  The axially-symmetric simulations were carried out for up to $2,000~\textrm{s}$ and resolved the radii down to $5\times 10^7~\textrm{cm}$ where the collapsing stellar material circularizes around the black hole.  The evolution of the central accretion rate in the simulations resembles the evolution of the observed GRB X-ray luminosity, which lends support to the hypothesis \citep{Kumar:08a,Kumar:08b} that the X-ray luminosity is proportional to the rate with which stellar material accretes onto the black hole.  We have identified three phases in the evolution of the accretion rate in our simulations, which appear to correspond to Phases 0 (the prompt phase), Phase I, and Phase II in the nomenclature of \citet{ZhangB:06}.

In the initial phase that in the simulations lasts $37-52~\textrm{s}$, the accretion of low-angular-momentum material is quasiradial for $r>5\times10^7~\textrm{cm}$ and occurs at quasi-constant rate of $\sim0.2~M_\odot~\textrm{s}^{-1}$.  The end of this phase is marked by the formation of an accretion shock at the smallest resolved radii.  The shock immediately propagates radially outwards through the supersonically infalling stellar envelope.  Simultaneously with the formation and the outer movement of the accretion shock, the accretion rate drops suddenly and precipitously.  We argue that the somewhat late onset of the accretion shock is an artifact of our not resolving the innermost two decades in radius outside the black hole's gravitational radius.

We supplement the simulations with an analytical model of the innermost accretion disk not resolved in the simulations, and suggest that the accretion shock forms early, within a fraction of the first second of the formation of the black hole, as several published simulations of the innermost neutrino-cooled region have shown, but only starts to propagate outward after $20~\textrm{s}$, when the material that is reaching the equatorial plane has enough angular momentum to circularize at radii where the virial temperature is below $\sim 10^{10}~\textrm{K}$ and the cooling by neutrino emission is suppressed.

During the second phase characterized by a steep decline $\propto t^{-2.7}$ of the accretion rate that lasts $\sim 500~\textrm{s}$, the accretion shock sweeps through the star, but a supersonic accretion of the shocked fluid in the axial funnel region proceeds unabated, at least in our simulations where the funnel has not been heated to high temperatures by the relativistic jet.  The thick disk containing rotationally-supported and pressure-supported fluid is convective; a high-entropy outflow from the inner, rotationally-supported region follows the accretion shock on its traversal through the star but remains bound within the star and appears to form a large-scale circulation pattern.  The steepness of the accretion rate decline seems to be the consequence of a rapid hydrodynamic readjustment of the shocked, convective, and circulating stellar envelope.  

The steep decline of the accretion rate slows down or stalls after $\sim 600~\textrm{s}$, which appears to reflect the settling of a fraction of the stellar envelope in the state of near-hydrostatic equilibrium.  The inner, rotationally supported thick disk contains $\sim 1\%$ of the mass of the unaccreted envelope and extends to $\sim 3\times10^9~\textrm{cm}$.  The thick disk is surrounded by a much more massive, \emph{pressure-supported} atmosphere, which acts as a mass supply to the thick disk. At no point do we find evidence for the extended thin disk envisioned by \citet{Cannizzo:09}.  The fluid above and below the thick disk is mostly unbound and the simulations thus exhibit a form of a ``disk wind.''

We speculate that depletion of the envelope through accretion onto the black hole or mass loss in thermal outflows or winds could be responsible for the renewed steepening of the GRB X-ray light curve after $10^3-10^4~\textrm{s}$.  More speculatively, the additional steepening of the light curve occasionally observed after $10^4-10^5~\textrm{s}$ could be due to a pervasive thermal or radiatively-driven mass loss in the outer layers of the atmosphere.

\acknowledgements

We would like to thank Ramesh Narayan and Craig Wheeler for encouraging this research at an early stage. We would also like to thank Stan Woosley and Alex Heger for making their pre-supernova stellar model 16TI available.  In addition, we would like thank Rodolfo Barniol Duran and Rongfeng Shen for valuable discussions.  The software used in this work was in part developed  by the DOE-supported ASC/Alliance Center for Astrophysical  Thermonuclear Flashes at the University of Chicago.   The authors acknowledge the Texas Advanced Computing Center  (TACC) at the University of Texas at Austin for providing high-performance computing resources that have contributed to this research.   This material is based upon work supported under a National Science Foundation Graduate Research Fellowship awarded to C.~C.~L.  M.~M. acknowledges support from NSF grant AST-0708795 and P.~K. acknowledges support from NSF grant AST-0909110.

\end{document}